\documentclass[aps,twocolumn,prd,superscriptaddress]{revtex4-2}

\usepackage[utf8]{inputenc}

\usepackage{mathtools}
\usepackage{amsfonts}
\usepackage{mathrsfs}
\usepackage{bbm}
\usepackage{physics}
\usepackage{slashed}
\usepackage{tensor}

\usepackage{graphicx}
\usepackage{color, float}
\usepackage{array}
\usepackage[abs]{overpic}

\usepackage{placeins}

\usepackage{makecell}
\usepackage{subcaption}
\usepackage{xspace}
\usepackage{siunitx}
\usepackage{xfrac}
\usepackage{hyperref}
\usepackage[nameinlink]{cleveref}
\usepackage{appendix}
\newcommand{\imag}{\text{i}}
\usepackage{xifthen}
\usepackage{xcolor}
\hypersetup{
	colorlinks,
	linkcolor={red!75!black},
	citecolor={blue!75!black},
	urlcolor={blue!75!black}
}

\usepackage{booktabs}
\usepackage{multirow}

\newcolumntype{C}{>{$}c<{$}}
\AtBeginDocument{
	\heavyrulewidth=.08em
	\lightrulewidth=.05em
	\cmidrulewidth=.03em
	\belowrulesep=.65ex
	\belowbottomsep=0pt
	\aboverulesep=.4ex
	\abovetopsep=0pt
	\cmidrulesep=\doublerulesep
	\cmidrulekern=.5em
	\defaultaddspace=.5em
}

\graphicspath{{./figures/}}

\newcommand{\gettitle}{Physics-informed gauge theories}

\newcommand{\getHeidelbergAffiliation}{\affiliation{Institut f{\"u}r Theoretische Physik, Universit{\"a}t Heidelberg, Philosophenweg 16, 69120 Heidelberg, Germany}}
\newcommand{\getEMMIAffiliation}{\affiliation{ExtreMe Matter Institute EMMI, GSI, Planckstr. 1, 64291 Darmstadt, Germany}}

\hypersetup{
	pdftitle={\gettitle},
	pdfauthor={Ihssen, Pawlowski},
	pdfkeywords= {renomalisation group} ,
	bookmarksopen=true,
	bookmarksopenlevel=2,
	bookmarksnumbered=true
}

\begin{document}
	
	\title{\gettitle}
	
	\author{Friederike Ihssen}
	%\thanks{ihssen@thphys.uni-heidelberg.de}
	\getHeidelbergAffiliation
	\author{Jan M. Pawlowski}\getHeidelbergAffiliation\getEMMIAffiliation
\begin{abstract}
We use the physics-informed renormalisation group (PIRG) for the construction of gauge invariant renormalisation group flows. 
The respective effective action is a sum of a gauge invariant quantum part and the classical gauge fixing part which arranges for invertibility of the gauge field two-point function. 
Thus, the BRST transformations simply accommodate the gauge consistency of the gauge fixing sector, while the quantum part of the effective action is gauge and BRST invariant. 
We apply this physics-informed approach to Yang-Mills theory and gravity and show how the flowing gauge fields arrange for full gauge invariance. 
We also embed the background field approximation to the functional renormalisation group (fRG) in an exact gauge invariant PIRG flow. This allows us to discuss the dynamics of the correction terms, and the non-trivial ultraviolet or infrared relevant terms are elucidated within a one-loop approximation. 
The background field  approximation of the latter is known for violating one-loop universality for specific regulators and we show how the present setup reinstates universality in a constructive way. 
Finally, we discuss the landscape of fRG flows in gauge theories through the lens of the novel PIRG approach as well as potential applications. 
\end{abstract}
	\maketitle

%%%%%%%%%%%%%%%%%%%%%%%%%%%%%%%%%
\section{Introduction} 
\label{sec:Introduction} 

In the present work we use the physics-informed renormalisation group (PIRG) approach \cite{Ihssen:2024ihp} for constructing gauge invariant RG flows. The construction builds on the key principle of PIRGs: flow equations including general reparametrisations are defined by the pair of the effective action, the \textit{target action}, and the corresponding emerging composite field, the \textit{flowing field}, see \cite{Ihssen:2023nqd}. Then, 
the generalised flow equation \cite{Pawlowski:2005xe} accommodates flows for both, the target action and the flowing field. It has been shown in \cite{Ihssen:2024ihp} as an extreme example that this even allows us to store all physics in the flowing field while keeping the effective action classical. In this case the task of solving diffusion-convection type partial differential equation already in simple approximations reduces to the task of solving simple first order ordinary differential equations. This huge computational advantage is paid for with a reconstruction problem for the correlation functions of the fundamental fields. However, this reconstruction problem has a systematic solution, see \cite{Ihssen:2024ihp}. 

It has been advocated in \cite{Ihssen:2024ihp} that PIRGs can be used to set up gauge invariant flows: we simply keep the \textit{quantum part} of the effective action gauge invariant and store the non-trivial part of the Slavnov-Taylor identities in the flowing gauge field. This gauge invariant PIRG is constructive and allows for gauge invariant flows in any approximation of the effective action. We use a background field approach as is natural for the gauge fixing in quantum gravity. Moreover, the reconstruction problem mentioned above for general PIRGs is avoided as the gauge invariant quantum part of the target action reduces to the standard background field effective action for vanishing flowing fields. Finally, the inherent flexibility and generality of the approach allows us to embed all fRG flows in gauge theories within the PIRG framework. Amongst possible applications of this embedding is the systematic improvement of a given approach, or to assess the systematic error of given approximations. 

In \Cref{sec:GaugeINvariantPIRGS} we set up the gauge invariant PIRG approach. Specifically we derive three core properties for this setup: a simple Nielsen identity (shift symmetry) \textbf{(i)}, gauge invariance \textbf{(ii)}, and BRST invariance \textbf{(iii)}, see \Cref{sec:FlowingGaugeInvariance}. The existence of flowing fields with these properties is proven in \Cref{sec:Flowingi-ii-iii}, and leads to simple gauge invariant flows and the respective disentangled flow for the flowing field in \Cref{sec:PIGsinNutshell}. In \Cref{sec:OneLoopPIGs} we illustrate some of the properties of the present gauge invariant flows, and discuss in particular the fate of relevant terms. In \Cref{sec:GaugeInvariantLandscape} we  provide a brief overview of fRG flows in gauge theories, embed them in the PIRG framework, and discuss possible practical applications. A short conclusion is provided in \Cref{sec:Conclusion}.

%%%%%%%%%%%%%%%%%%%%%%%%%%%%%%%%%%
\section{Gauge invariant PIRGs} 
\label{sec:GaugeINvariantPIRGS}

The physics-informed RG approach set up in \cite{Ihssen:2024ihp} accommodates general renormalisation group flows including general reparametrisations of the theory. It is the latter property which is exploited for setting up gauge invariant and gauge-consistent (BRST-invariant) fRG flows.

%%%%%%%%%%%%%%%%%%%%%%%%%%%%%%%%%%
\subsection{PIRGs and their properties} 
\label{sec:PIRGProperties}

The generality of the PIRGs is obtained by formulating the functional renormalisation group for the pair  
\begin{align} 
\bigl( \Gamma_\phi[\phi]\,,\,\dot\phi[\phi]\bigl)\,,
\label{eq:PairGamm-phi}
\end{align}
where $\Gamma_\phi[\phi]$ is the \textit{target action}. The field $\phi$ is the \textit{flowing field} or emergent composite. While it is the variable of $\Gamma_\phi$, it can also be understood as a functional of the fundamental field $\varphi$ of the theory. Its change with the RG-scale $k$ is given by $\dot\phi[\phi]$, which can be defined to accommodate a specific purpose. In terms of the field operators $\hat \varphi, \hat\phi$ this entails 
\begin{align}
	\phi =\langle \hat \phi_k[\hat\varphi]\rangle \,, \qquad \mathrm{and} \qquad \dot\phi[\phi]= \langle \partial_t \hat \phi_k[\hat\varphi]\rangle \,,
	\label{eq:phihatphi}
\end{align}
where we allow an RG-scale dependence of the transformation and the RG-time is given by $t=\log k/k_\textrm{ref}$ with some reference scale $k_\textrm{ref}$. The composite field $\hat\phi$ is coupled to a current $J_\phi$ in the path integral. Consequently, the Legendre transform leading to the effective action $\Gamma_\phi$ is performed with respect to the current $J_\phi$. 

The important change of perspective advertised in \cite{Ihssen:2024ihp} is the following: The mean field $\phi$ is the independent variable of the effective action  $\Gamma_\phi[\phi]$ and carries no $k$-dependence. Moreover, $\dot \phi[\phi]$ is a functional of $\phi$ and is at our disposal. This flexibility can be used to construct a wide variety of target actions. Limitations have been discussed judiciously in \cite{Ihssen:2024ihp}. The  construction entails that in general, $\Gamma_\phi$ is not the effective action of the fundamental field, $\Gamma_\varphi$, rewritten in terms of the flowing field but a different generating functional. Still, it carries the same physics information as $\Gamma_\varphi$. Notably, we do not need to know the underlying transformation $\partial_t \hat\phi_k[\hat\varphi]$, \labelcref{eq:phihatphi},  or rather $\partial_t \hat\phi_k[\hat\phi]$ for this setup to reconstruct physical observables, since correlation functions of the fundamental field can be recovered directly from $\Gamma_\phi$. For a detailed discussion of the reconstruction process see \cite{Ihssen:2024ihp}. 
Moreover, the reconstruction is not necessary in the present PIRG approach to gauge theory, since the gauge invariant action reduces to the standard background field effective action for vanishing flowing field. This will be discussed in \Cref{sec:PIRGsforPIGs}. 

The pair \labelcref{eq:PairGamm-phi} of a target action and the respective flowing field satisfies the generalised flow equation \cite{Pawlowski:2005xe}, 
\begin{align}\nonumber 
\left( \partial_t + \int_x \dot{\phi} [\phi] \frac{\delta}{\delta \phi} \right) \Gamma_\phi[\phi] & \\[1ex]
&\hspace{-1.4cm}=\frac{1}{2} \Tr \left[G_{\phi\phi}[\phi]\left(\partial_t + 2 \frac{\delta \dot\phi }{\delta \phi}  \right) \, R_k \right]\,, 
	\label{eq:GenFlow} 
\end{align}
with the propagator $G_{\phi\phi}$ of the flowing field $\phi$, 
\begin{align} 
	G_{\phi\phi}[\phi] =\frac{1}{\Gamma_\phi^{(2)}[\phi]+R_k}\,,\qquad \Gamma_{\phi}^{(n)}[\phi]=\frac{\delta^n \Gamma_\phi}{\delta \phi^n}\,. 
	\label{eq:Gphiphi}
\end{align}
This completes the PIRG setup. In short, it allows to use the pair \labelcref{eq:PairGamm-phi} in combination with the generalised flow \labelcref{eq:GenFlow} to simplify the conceptual and numerical tasks at hand. 

We close this introduction with some remarks on the potential and the restrictions of the approach. We concentrate on the properties relevant for the gauge invariant PIRG setup and refer for a more detailed discussion to \cite{Ihssen:2024ihp}. The existence of the pair \labelcref{eq:PairGamm-phi} for a given target action is specially important for the construction of gauge invariant PIRGs. Indeed, the gauge invariant target action has to satisfy three properties already mentioned in the introduction: a simple Nielsen identity (shift symmetry) \textbf{(i)}, gauge invariance \textbf{(ii)}, and BRST invariance \textbf{(iii)}. They will be discussed later in \Cref{sec:FlowingGaugeInvariance}. These properties require general reparametrisations $\dot\phi$. Below we discuss a few construction principles and their limitations:  
\begin{itemize}
\item[\textbf{(a)}] Interpreting the generalised flow equation as a constraint equation for the flowing field $\dot \phi[\phi]$, allows for an iterative construction of the field transformation for a given target action $\Gamma_\phi$ in orders of perturbation theory: the $n$-loop order contribution to $\dot\phi$ on the left hand side of \labelcref{eq:GenFlow} can be computed by the $n$-loop order contribution of the right-hand side. The latter is a loop itself and its computation only needs $(n-1)$-loop orders of $\dot\phi$ (and $\Gamma_\phi$). Hence, a perturbation theory solution can be constructed iteratively at all orders. 

\item[\textbf{(b)}] For a given approximation of the effective action with a finite number of operators, a solution $\dot\phi$ can be constructed by projecting \labelcref{eq:GenFlow} on these terms and solving the finite number of constraint equations. This includes approximations with full effective potentials and fully momentum-dependent vertices. This encompasses all approximations used in the literature. 

\item[\textbf{(c)}] The existence of the target action at $k=0$, derived from the pair \labelcref{eq:PairGamm-phi}, is ensured by integrating the flow of $\dot\phi$: If this flow is non-singular, the target action and hence the reparametrisation exists.   
\end{itemize} 
In summary, in the validity regime of perturbation theory, this construction can readily be applied for general reparametrisations. Moreover, all approximations and approximation schemes used in fRG computations in gauge theories and beyond, are covered by \textbf{(b)}. This also includes the 'radical' choice of a classical target action, $\Gamma_\phi[\phi] = S_\textrm{cl}[\phi]$ with the classical action $S_\textrm{cl}$. The latter case has been successfully tested in \cite{Ihssen:2024ihp} within zero dimensions, where the effective potential is the exact effective action. 

Finally, the global existence of the reparametrisation follows from the non-trivial constraint \textbf{(c)}. Its validity is specifically interesting for the global existence of gauge invariant flows: It is relevant both for the access and stability of ultraviolet fixed points in quantum gravity within gauge invariant flows, as well as the existence of global ultraviolet-infrared trajectories. In short, the PIRG setup allows to elevate approximate flow equations to exact ones if the solution of the approximate flow is interpreted as a target action $\Gamma_T= \Gamma_\phi$, and \labelcref{eq:GenFlow} is then seen as a constraint equation for the flowing field. This allows us to elevate approximate flows to consistent ones, while not being complete. Here we use the classification put forward in  \cite{Litim:2002xm}: A \textit{complete} flow is one that generates the \textit{complete} dynamics of the theory under investigation, while a \textit{consistent} flow is an exact one, where possibly not the full dynamics is generated by the flow of the effective action. 

\begin{figure}[t]
	\centering
	\includegraphics[width=0.36\columnwidth]{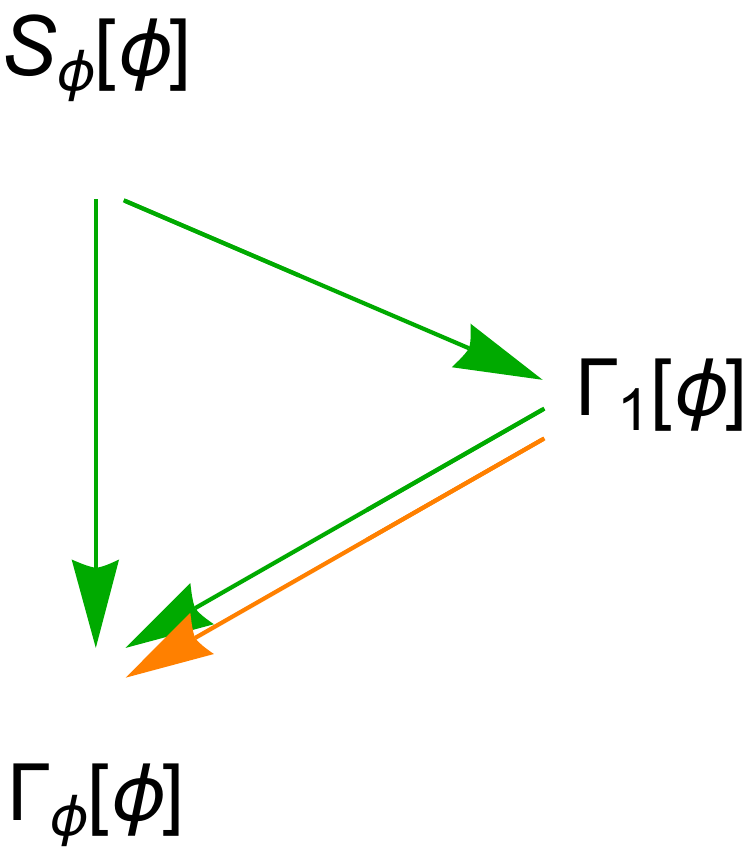}\\[1.33ex]
	\caption{ Complete (green) and consistent (orange) fRG flows for the effective action of the fundamental field.  \hspace*{\fill} }
	\label{fig:CompleteConsistentGphif}
\end{figure}

The original classification is illustrated in  \Cref{fig:CompleteConsistentGphif}. A complete flow (green) connects the classical or fixed point action for $k\to\infty$ with the full effective action at $k=0$. A consistent flow (orange) connects a well-defined initial condition $\Gamma_1$ at $k=\Lambda$ with the full effective action, or another well-defined generating functional, at $k=0$. A simple non-trivial example for a consistent but incomplete flow are thermal flows that only integrate out thermal fluctuations. They lead from the vacuum theory $\Gamma_1=\Gamma_{T=0}$ to the thermal one, $\Gamma=\Gamma_T$, see \cite{Litim:2002xm}. Evidently, complete flows are also consistent.

\begin{figure}[t]
	\centering
	\parbox{0.5\linewidth}{\includegraphics[width=0.25\columnwidth]{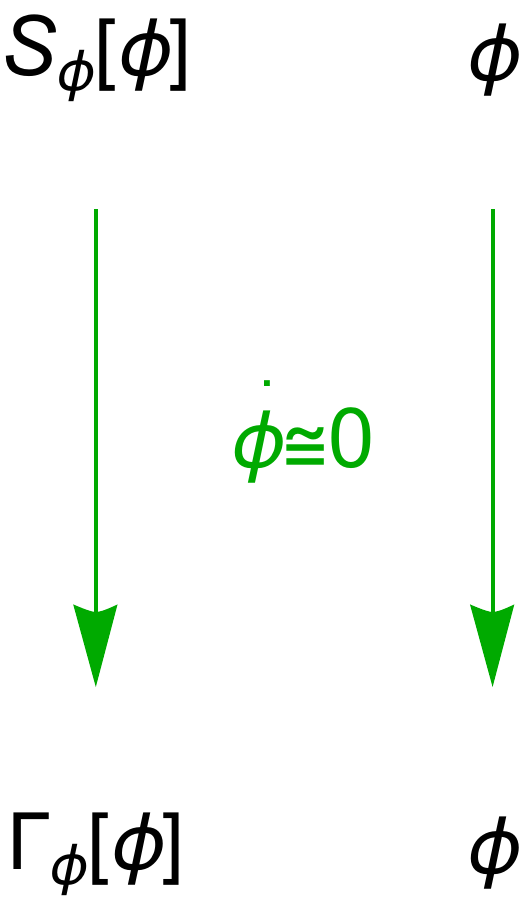}
	}%
	\parbox{0.5\linewidth}{\includegraphics[width=0.25\columnwidth]{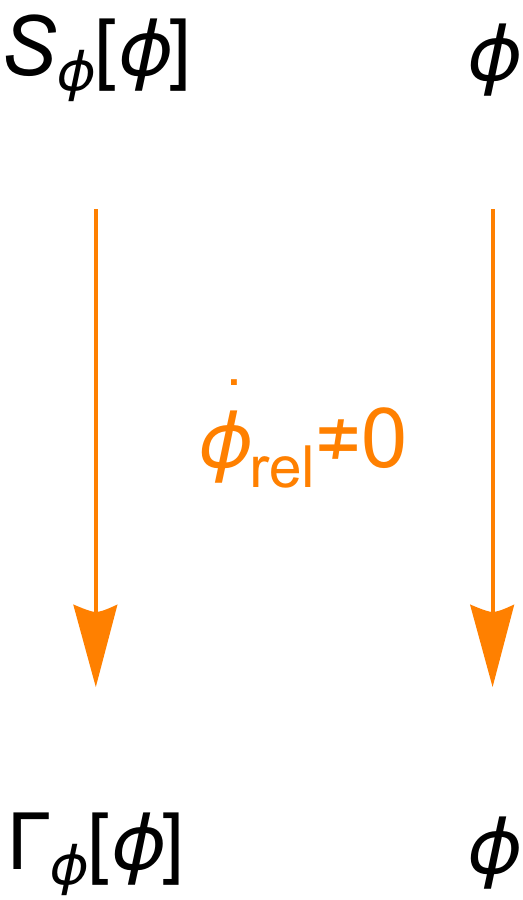}
	}
	\caption{Complete (green, left) and consistent (orange, right) PIRG flows for the pair of the target action $\Gamma_\phi$ and the flowing field.  \hspace*{\fill} }
	\label{fig:CompleteConsistentGphi}
\end{figure}

Now we extend the classification scheme to PIRGs: we call a flow complete if the target action contains the complete dynamics of the theory. We emphasise that this does not imply that the flowing field vanishes, as it may only contain a reparametrisation of the path integral. Within this classification scheme, PIRGs are certainly consistent: The target action and its flow $\partial_t\Gamma_\phi$ comprise part of the dynamics, while the flowing field $\dot \phi$ carries the rest as \labelcref{eq:GenFlow} is exact. If the flowing field only carries a simple rescaling of the fundamental field, the flow $\partial_t \Gamma_\phi$ is complete (green), see left plot in \Cref{fig:CompleteConsistentGphi}. In general, the flowing field carries part of the dynamics, and in particular it carries a part of the ultraviolet or infrared relevant operators. Such a flow is incomplete, but consistent (orange), see right plot in \Cref{fig:CompleteConsistentGphi}. 
Here, we indicated the incompleteness with $\left. \dot\phi\right|_\textrm{rel}\neq 0$. Note however that even with a vanishing relevant part of the flowing field, PIRG flows are in general incomplete.  
This classification is important for the assessment of the background field approximation in gauge theories or general proper time flows in \Cref{sec:OneLoopPIGs}.

%%%%%%%%%%%%%%%%%%%%%%%%%%%%%
\subsection{Physics-informed gauge theories}
\label{sec:PIRGsforPIGs}

With these preparations we formulate PIRG-flows for general gauge theories, using non-Abelian gauge theories with the gauge field $A_\mu$ and metric quantum gravity with the gauge field $g_{\mu\nu}$ as explicit examples. The pair \labelcref{eq:PairGamm-phi} in a gauge theory is given by 
\begin{align} 
	\bigl( \Gamma_\phi[\bar\varphi,\Phi]\,,\,\dot\Phi[\bar\varphi,\Phi]\bigl)\,, 
	\label{eq:PairGamm-phiGauge}
\end{align}
where $\Phi$ is a superfield that contains the flowing gauge field $\phi$ and potential auxiliary fields such as ghosts and the Nakanishi-Lautrup field, see \labelcref{eq:FlowingPhi}. Below we discuss its definition and use in the present approach. Importantly, the flowing gauge field is a general functional of the fundamental fields of the theory at hand. Finally, the field $\bar\varphi$ in \labelcref{eq:PairGamm-phiGauge} is a background gauge field. 

A first non-trivial difference to the standard approach is that we discuss the gauge fixing sector on the level of the effective action $\Gamma_\phi$ with the flowing mean gauge field $\phi$, defined by its transformation $\dot \phi[\phi]$. In the standard approach the gauge fixing is introduced on the level of the classical action with the microscopic field $\hat A_\mu$ or $\hat g_{\mu\nu}$, see \Cref{app:BackgroundYM} and \Cref{app:BackgroundGravity} respectively. The respective mean fields are given by 
\begin{align} 
	A_\mu = \langle \hat A_\mu\rangle\,,\qquad \mathrm{and} \qquad  g_{\mu\nu} = \langle \hat g_{\mu\nu}\rangle\,. 
\end{align}
In both example theories we resort to the background field approach. The standard background approach splits the full gauge field operator $\hat \varphi$ with $\hat\varphi=\hat A_\mu$ or $\hat \varphi=\hat g_{\mu\nu}$ into a background field and a fluctuation field,  
\begin{align}
	\hat\varphi = \bar \varphi + \Delta \hat\varphi\,,\qquad \Delta \hat\varphi_\mu = \hat a_\mu\,,\qquad \Delta \hat\varphi_{\mu\nu} = \hat h_{\mu\nu}\,.
	\label{eq:StandarLinearSplit}
\end{align}
With this \textit{linear} split of the fundamental field the mean gauge fields are given by 
\begin{align} 
	\varphi=\bar\varphi+ \Delta \varphi\,.
	\end{align}
Finally, the Yang-Mills and gravity fluctuation mean fields $\Delta\varphi= \langle \Delta\hat\varphi \rangle $ with $\Delta\varphi_\mu= a_\mu$ (Yang-Mills) and  $\Delta\varphi_{\mu\nu}= h_{\mu\nu}$ (gravity) are given by 
\begin{align} 
 a_\mu = A_\mu - \bar A_\mu\,,\qquad \mathrm{and} \qquad h_{\mu\nu} =g_{\mu\nu}  -\bar g_{\mu\nu} \,.
\end{align}
Note also that in gravity one typically also uses a computationally convenient prefactor $\sqrt{G_N}$ in front of the fluctuation field to obtain the canonical momentum dimension $1$ for the fluctuation field. In the present work we refrain from using this as we are more interested in conceptual questions. 

In the present PIRG approach, we only discuss mean fields. Hence, we introduce a mean field $\chi$ as a linear split of the background gauge field and the flowing fluctuation field $\phi$,   
\begin{align}
	\chi= \bar \varphi + \phi\,, 
	\label{eq:LinearFlowingSplit}
\end{align}
with the explicit splits 
\begin{align}
\chi_\mu=\bar A_\mu + \phi_\mu\,,\qquad \mathrm{and} \qquad \chi_{\mu\nu} = \bar g_{\mu\nu} + \phi_{\mu\nu}\,.
	\label{eq:chi}
\end{align}
Importantly, the fluctuation gauge fields $\phi_\mu$ and $\phi_{\mu\nu}$ are \textit{not} the fundamental fluctuation fields $a_\mu$ and $h_{\mu\nu}$, but composite operators which remain to be specified. We introduce them as general functions of the fundamental mean fields, $\phi=\phi[\bar \varphi,\Delta \varphi]$, where we have dropped the dependences on the auxiliary fields. The respective field operator is given by 
\begin{align}
\hat\chi  =\bar\varphi +\hat\phi_k[\bar\varphi,\hat\varphi]\,, 
\label{eq:hatomega}
\end{align}
which can also be expressed in terms of $\bar\varphi$ and $\Delta\hat\varphi$. We emphasise that the approach does not make use of the operator relation \labelcref{eq:hatomega}, but only its mean field analogue \labelcref{eq:chi}. Furthermore, $\varphi$ and $\chi$ are different objects. However, they reduce to the gauge field background for $\Delta \varphi=0$ and $\phi=0$.

%%%%%%%%%%%%%%%%%%%%%%%%%%%%%%
\subsection{Flowing gauge and BRST invariance}
\label{sec:FlowingGaugeInvariance} 

The flowing gauge fields $\phi$ are general functions of the fields $A_\mu$ or $g_{\mu\nu}$ as well as the standard auxiliary fields within a given gauge fixing, the ghosts $c,\bar c$ and potentially a Nakanishi-Lautrup field $b$ for full BRST invariance. In the following we only discuss the standard covariant background gauge fixing as an example. While the gauge invariant PIRG approach is general, we consider this gauge fixing a very convenient and practical choice. The background gauge fixing sector is given by 
\begin{align}
	S_\textrm{gauge}[\bar \varphi,\Phi] = S_\textrm{gf}[\bar \varphi,\Phi] + S_\textrm{gh}[\bar \varphi,\Phi]\,, 
	\label{eq:GaugeSector}
\end{align}
with the flowing superfield 
\begin{align} 
	\Phi=(\phi, c, \bar c, b)\,,\qquad \bar\varphi +\Phi = (\bar \varphi+\phi, c, \bar c, b)\,.
	\label{eq:FlowingPhi}
\end{align}
Here we have suppressed Lorenz and group indices to keep the discussion general. The gauge fixing sector in our two example theories is discussed in \Cref{app:BackgroundYM} and \Cref{app:BackgroundGravity}. For Yang-Mills theory it is given by \labelcref{sec:ActionYMBack}, for metric gravity by \labelcref{eq:SgaugeGravity}. As we use Yang-Mills theory for the explicit computations here, we also provide details on the standard background field fRG approach to Yang-Mills theories in \Cref{app:BackgroundYM}. In particular this includes the modified master equation  \labelcref{eq:mMaster} which encodes the breaking of gauge invariance, and the modified Nielsen identity \labelcref{eq:modNielsenID}. 

With these preparations we proceed with the completion of our PIRG setup with a gauge invariant target action $\Gamma_\phi$ within a gauge-fixed theory. This requires a flowing field which is potentially a non-trivial functional of the fundamental fields, even for large cutoff scales. Hence, the gauge fixing \labelcref{eq:GaugeSector} with \labelcref{sec:ActionYMBack,eq:SgaugeGravity} is only seemingly linear. The demanded gauge invariance of $\Gamma_\phi$ is implemented by a flowing field setup for physics-informed gauge theories with the following three properties \textbf{(i,ii,iii)}:

\textbf{(i)} \textit{Trivial Nielsen identity:} The quantum part of the target action only depends on the sum of the background gauge field and the flowing fluctuation gauge field,    
\begin{subequations} 
	\label{eq:PIRGsforPIGs}
\begin{align} 
	\Gamma_\phi[\bar \varphi, \Phi]= \Gamma_{\phi,\textrm{qu}}[\bar\varphi +\Phi] + S_\textrm{gauge}[\bar \varphi,\Phi]\,,
	\label{eq:Property-i}
\end{align}
with the 'classical' gauge fixing sector \labelcref{eq:GaugeSector} in the standard background field gauge. \Cref{eq:Property-i} is the Nielsen identity and encodes the shift symmetry in the gauge field. It is commonly written in its derivative form, 
\begin{align} 
	\left(\frac{\delta}{\delta  \phi}- \frac{\delta}{\delta \bar \varphi} \right) \Gamma_{\phi,\textrm{qu}}[\bar\varphi, \Phi]=0\,. 
	\label{eq:Derback=Derfluc}
\end{align}
The standard Nielsen identity in an fRG setup is discussed in \Cref{app:Nielsen}. \\[-1ex] 
	
\textbf{(ii)} \textit{Background gauge invariance:} The target action is invariant under background gauge transformations,  
\begin{align}
	\bar \delta_\omega \Gamma_\phi[\bar\varphi,\Phi]=0\,. 
	\label{eq:Property-ii}
\end{align}
Background gauge invariance and the respective transformation are discussed in detail in \Cref{app:BGgauge}. \\[-1ex] 

\textbf{(iii)} \textit{BRST invariance:} The target action is BRST-invariant, i.e.~gauge-consistent. In the presence of the Nakanishi-Lautrup field it satisfies the master equation  
\begin{align}
 \frac{\delta \Gamma_\phi}{\delta \mathcal{Q}_a}  \frac{\delta \Gamma_\phi}{\delta \Phi_a}
	=0 
	\, . 
\label{eq:Property-iii}
\end{align}
\end{subequations} 
In \labelcref{eq:Property-iii} we have used DeWitt's condensed notation and the index $a$ now also includes an integral over space-time. The $Q_i$ are the currents for the BRST transformations of the field $\Phi_i$. The BRST setup including the standard modified master equation is discussed in \Cref{eq:BRST+Gauge}.\\[-1ex]

PIRGs in gauge theories with the properties \textbf{(i) - (iii)} are gauge and BRST invariant. While BRST invariance is the property \textbf{(iii)}, physical 
gauge invariance of $\Gamma_{\phi,\textrm{qu}}$ follows readily from background gauge invariance \labelcref{eq:Property-ii} and the trivial Nielsen identity \labelcref{eq:Derback=Derfluc}: Since $\Gamma_{\phi,\textrm{qu}}= \Gamma_{\phi,\textrm{qu}}[\bar\varphi+\Phi]$, the gauge field part of the background gauge transformation $\bar\delta_\omega$ is the quantum gauge transformation $\delta_\omega$. Augmenting $\delta_\omega$ in \labelcref{eq:GaugeTrafos} with the background transformations of $c,\bar c,b$ leads to 
\begin{align} 
	\delta_\omega \Gamma_{\phi,\textrm{qu}}[\bar\varphi+\Phi]=0\,.
\label{eq:GaugeInvariantPIG}
\end{align} 
\Cref{eq:GaugeInvariantPIG} expresses the desired physical gauge invariance. 

We complement this discussions with a few remarks.  We start with brief interpretations of the properties. \Cref{eq:Property-i} implements a trivial version of the Nielsen identity. \Cref{eq:Property-iii} is the most challenging property: it requires that the non-trivial loop terms of the modified Master equation \labelcref{eq:mMaster} are absorbed in the flowing gauge field. Note also that, if integrating out the Nakanishi-Lautrup field $b$, \labelcref{eq:Property-iii} turns into the standard Zinn-Justin equation.

Furthermore, these properties are not independent from each other. For example, property \textbf{(ii)} follows readily from \textbf{(i)} and \textbf{(iii)}. The latter seemingly arranges for a fully gauge invariant covariant momentum cutoff. This has to be taken with a grain of salt as the regulators are defined in terms of the background Laplacian. \Cref{eq:Property-iii} is achieved by a flowing gauge field transformation whose inherent non-linearity in terms of the fundamental field makes the underlying BRST transformations non-linear as well. This is an obstacle for the co-homological interpretation of the construction, for related considerations for supersymmetry and chiral symmetry see \cite{Bergner:2012nu}. 

The setup is completed with the generalised flow equation for the target action which is the gauge theory version of \labelcref{eq:GenFlow}. This flow guarantees that the three properties \labelcref{eq:PIRGsforPIGs} hold true at all cutoff scales, including $k=0$. 
The respective generalised flow equation for the gauge invariant effective action reads 
\begin{subequations} 
\label{eq:GenGaugeFlows}
\begin{align}\nonumber 
	\partial_t \Gamma_{\phi,\textrm{qu}}[\bar \varphi+ \Phi] = &-   \dot{\Phi}_a [\bar \varphi, \Phi] \frac{\delta \Gamma_{\phi}[\bar \varphi,\Phi]}{\delta \Phi_a}  \\[1ex] 
	&\hspace{-2.3cm}+\frac{1}{2} G_{ab}[\bar \varphi, \Phi]\left(\partial_t \delta^b_c + 2 \frac{\delta \dot{\Phi}_c [\bar \varphi, \Phi] }{\delta \Phi_b}\right) \, R_k^{ac}[\bar \varphi] \,, 
	\label{eq:GenGaugeFlowEquation} 
\end{align}
with the full field-dependent propagator of the flowing fluctuation fields 
\begin{align} 
	G[\bar \varphi,\Phi] =\frac{1}{\Gamma_{\phi,\textrm{qu}}^{(0,2)}[\bar \varphi+\Phi]+S_\textrm{gauge}^{(0,2)}[\bar\varphi,\Phi]+ R_k[\bar \varphi]}\,, 
	\label{eq:GPhhiPhi}
	\end{align} 
and the mixed background and fluctuation derivatives of the effective action 
\begin{align}\nonumber 
	\Gamma_\phi^{(n,m)}[\bar \varphi,\Phi]=&\, \frac{\delta^{n+m}\Gamma_\phi[\bar \varphi,\Phi]}{\delta \bar\varphi^n\delta\Phi^m} \\[1ex]
	=&\, \Gamma_{\phi,\textrm{qu}}^{(0,n+m)}[\bar \varphi+\Phi]+S_\textrm{gauge}^{(n,m)}[\bar\varphi,\Phi]\,.
	\label{eq:Gnm}
\end{align}
\end{subequations}
In comparison to \labelcref{eq:GenFlow} we have re-ordered 	\labelcref{eq:GenGaugeFlowEquation}  such that the left hand side only contains the $t$-derivative of the gauge invariant quantum part of the target action.

%%%%%%%%%%%%%%%%%%%%%%%%%%%%
\section{Flowing fields for gauge invariant PIRGs}
\label{sec:Flowingi-ii-iii}

In this Section we discuss, whether the three properties \textbf{(i,ii,iii)} can be implemented with the generalised flows. To that end we will assume that all properties are satisfied at some scale $k$, and show that a flowing field $\dot \Phi$ can be found for which the flow $\partial_t \Gamma_\phi$ has all three properties. For such a proof it is important to realise that the transformation $\dot\Phi$ is very powerful. Indeed, as discussed in \Cref{sec:PIRGProperties}, if the approximation of the effective action only contains a finite number or operators, it can even be used to absorb the flow completely into the transformation. This would reduce \labelcref{eq:GenGaugeFlowEquation} to 
\begin{align}\nonumber 
		\partial_t \Gamma_{\phi,\textrm{qu}}= & 0 \,,  \\[1ex] 
  \dot{\Phi}_a  \frac{\delta \Gamma_{\phi}}{\delta \Phi_a}	=& \frac{1}{2} G_{ab}\,\left(\partial_t \delta^b_c + 2 \frac{\delta \dot{\Phi}_c }{\delta \Phi_b}\right) \, R_k^{ac} \,, 
\end{align}
the classical target action flow. Trivially, the classical target action flow has all the properties \textbf{(i,ii,iii)}. In \cite{Ihssen:2024ihp} the existence of this solution has been exemplified with a very simple computational showcase, the zero-dimensional case, where the 'path integral' can be solved analytically. We emphasise that in this case the classical target action flow was reproducing the correct result not for the triviality of the example, but for the \textit{finite} number of operators involved: in zero dimensions the effective potential is the full effective action.

In the present case we do not want to store the physics or dynamics of a gauge theory in the transformation, but we only want to use the flowing field to absorb \textit{corrections} to either gauge symmetry relations, that are the properties \textbf{(ii,iii)}, or the Nielsen identity that relates them, that is property \textbf{(i)}. Nonetheless, the example illustrates that all three properties can be solved together. Therefore, the following analysis only details the nature of the fluctuations that are stored in the flowing field $\dot\Phi$.

%%%%%%%%%%%%%%%%%%%%%%%%%%%
\subsection{Property \textbf{(i)}: Trivial Nielsen identity}
\label{sec:Flowingi}

We start this analysis with the trivial Nielsen identity \labelcref{eq:Derback=Derfluc}. Its flow is given by  
\begin{align} 
	\left(\frac{\delta}{\delta  \phi}- \frac{\delta}{\delta \bar \varphi} \right) \partial_t \Gamma_{\phi,\textrm{qu}}[\bar\varphi,\Phi]=0\,, 
	\label{eq:FlowProperty-i}
\end{align}
where we have used that $\partial_t \Gamma_{\phi}= \partial_t \Gamma_{\phi,\textrm{qu}}$. 
Instead of aiming at the classical target action solution which trivially solves \labelcref{eq:FlowProperty-i}, we aim at the converse, the \textit{minimal} flowing field $\dot\Phi$ solving \labelcref{eq:FlowProperty-i}. To that end we remind ourselves that for $\dot\Phi=0$, the full dynamics of the theory is contained in the loop term 
\begin{align} 
	\textrm{Flow}[\bar\varphi,\Phi]= \frac12  G_{ab}[\bar\varphi,\Phi]\, \partial_t R^{ab}[\bar\varphi]\,.
\end{align}
Indeed, this term sources the different dependence on $\bar\varphi$ and $\phi$ in the flow: it depends on the following operators, 
\begin{align} 
	\Gamma^{(0,2)}_{\phi,\textrm{qu}}[\bar \varphi+\Phi]\,, \quad S_\textrm{gauge}^{(0,2)}[\bar\varphi,\bar \varphi+\Phi]\,,\qquad R_k[\bar\varphi] \,.
\label{eq:OpsWetterich} 
\end{align} 
While $\Gamma^{(0,2)}$ only depends on the sum $\bar \varphi+\Phi$, the other two operators depend on $\bar\varphi$ and $\Phi$ separately. For later convenience, we have written the operator $S_\textrm{gauge}^{(0,2)}$ in \labelcref{eq:OpsWetterich} as a function of the full gauge field $\bar\varphi+\phi$ and its dependence on $\bar\varphi$ instead of as a function of $\bar\varphi$ and $\phi$. 

There are two natural ways to guarantee \labelcref{eq:FlowProperty-i} for the flow of the Nielsen identity: We may simply drop $\bar\varphi$ in $S_\textrm{gauge}^{(2)}$ and $R_k$. Then, the gauge fixing reduces to the standard covariant gauge fixing and the regularisation reduces to one in momenta instead of covariant momenta. This formulation is the one commonly used in QCD, see the reviews \cite{Gies:2006wv, Braun:2011pp, Dupuis:2020fhh, Fu:2022gou}, as well as in the fluctuation approach to gravity, see the reviews \cite{Dupuis:2020fhh, Pawlowski:2020qer, Pawlowski:2023gym}.  

The second natural possibility lifts the background dependence on $\bar\varphi$ in $S_\textrm{gauge}^{(2)}$ and $R_k$ to that of the full field,
\begin{align} \nonumber 
S_\textrm{gauge}^{(0,2)}[\bar\varphi,\bar\varphi+\Phi]\ \to &\  {\cal D}[\bar\varphi+\Phi]=S_\textrm{gauge}^{(0,2)}[\bar\varphi+\phi, \bar\varphi+\Phi]\,,\\[1ex]
R_k[\bar\varphi]\ \to& \  R_k[\bar\varphi+\phi]\,.
	\label{eq:Backgrounddynamics} 
\end{align} 
This is done in the standard background field approximation to non-Abelian gauge theories and gravity. In particular, \labelcref{eq:Backgrounddynamics} introduces the dynamical background field into the regulator. If using such a regulator in the derivation of the flow equation, it leads to a flow equation with a series of $n$-loop terms instead of the one-loop exact one. It has been shown in \cite{Litim:2002ce} that within this approximation even one-loop universality is lost. Moreover, in gravity the background field approximation leads to qualitative changes in the flows, with severe consequences for the existence of fixed points in particular for matter-gravity systems \cite{Meibohm:2015twa}. For a detailed discussion see the review \cite{Pawlowski:2020qer}. 

In the present approach we rectify this deficiency by accommodating this change in the flowing field, thus re-instating an exact flow. To that end we rewrite the flow term as 
\begin{subequations}
\begin{align}
	\textrm{Flow}[\bar\varphi,\Phi]= 	\textrm{Flow}[\bar\varphi+\Phi] + \Delta\textrm{Flow}[\bar\varphi,\Phi]\,,
	\label{eq:DefFlow}
\end{align}
with
\begin{align}  
	\textrm{Flow}[\bar\varphi+\Phi] = &\,  \frac12  G_{ab}[\bar\varphi+\Phi]\, \partial_t R^{ab}[\bar\varphi+\phi]\,, 
\label{eq:FlowExplicit}
\end{align}
\end{subequations}
and the propagator 
\begin{align} 
G[\bar\varphi+\Phi] = \frac{1}{\Gamma^{(0,2)}_{\phi,\textrm{qu}}[\bar\varphi+\Phi]+{\cal D}[\bar\varphi+\Phi]+R_k[\bar\varphi+\phi]}\,,  
\label{eq:Gvarphi+Phi}
\end{align}
and ${\cal D}[\bar\varphi+\Phi]=S_\textrm{gauge}^{(0,2)}[\bar\varphi+\phi, \bar\varphi+\Phi]$, see \labelcref{eq:Backgrounddynamics}. Now we insert this reparametrisation into the flow of the trivial Nielsen identity \labelcref{eq:FlowProperty-i}. Using the generalised flow \labelcref{eq:GenGaugeFlowEquation}, we arrive at the constraint equation for $\dot\Phi$,  
\begin{align}\nonumber 
	\left(\frac{\delta}{\delta  \phi}- \frac{\delta}{\delta \bar \varphi} \right) \left[ \dot{\Phi}_a \frac{\delta \Gamma_\phi}{\delta  \Phi_a}  - G_{ab}\,\frac{\delta \dot{\Phi}_c }{\delta \Phi_b} \, R_k^{ac}\right]&\\[1ex]
	&\hspace{-3cm}=	\left(\frac{\delta}{\delta  \phi}- \frac{\delta}{\delta \bar \varphi} \right) \Delta\textrm{Flow}[\bar\varphi,\Phi] \,. 
	\label{eq:FlowNielsen}
\end{align} 
The expression in the square bracket on the left hand side defines the flowing field and has not been specified yet. The solution to \labelcref{eq:FlowNielsen} is given by 
\begin{align} 
 \dot{\Phi}_a \frac{\delta \Gamma_\phi}{\delta  \Phi_a}  - G_{ab}\,\frac{\delta \dot{\Phi}_c }{\delta \Phi_b} \, R_k^{ac}
= \Delta\textrm{Flow} -F_\phi[\bar\varphi+\Phi]\,, 
	\label{eq:FlowNielsenSolution}
\end{align}
with a general solution $F_\phi$ of the homogeneous differential equation (vanishing right-hand side of \labelcref{eq:FlowNielsen}). We may need a specific $F_\phi[\bar\varphi+\Phi]$ on the right-hand side to arrange for property \textbf{(iii)}, see \Cref{sec:Flowingiii}. In the discussion below we assume that it is either vanishing or a small correction. We will come back to its definition and verify the assumption in \Cref{sec:Flowingiii}. 

We close this Section with a short discussion of the consequences of \labelcref{eq:FlowNielsenSolution}, and of its nature as a small/minimal flowing field arranging for \labelcref{eq:Property-i}. We first insert \labelcref{eq:FlowNielsenSolution} into the generalised flow equation \labelcref{eq:GenGaugeFlowEquation}. This leads us to 
\begin{align} 
	\partial_t \Gamma_{\phi,\textrm{qu}}[\bar\varphi+\Phi] = \textrm{Flow}[\bar\varphi+\Phi]+F_\phi[\bar\varphi+\Phi]\,, 
\label{eq:FlowGqu}
\end{align} 
where $\textrm{Flow}[\bar\varphi+\Phi]$ is defined in \labelcref{eq:FlowExplicit}. For $F_\phi=0$, the flow \labelcref{eq:FlowGqu} is nothing but the flow in the background field approximation. The present derivation embeds this flow in an exact PIRG approach: While it is not an exact flow in terms of the linear split of the fundamental gauge fields, \labelcref{eq:StandarLinearSplit}, it is an exact flow in terms of the non-trivial flowing gauge field $\phi$ with \labelcref{eq:FlowNielsenSolution}. 

It is worth emphasising that the pair or rather the split $(\Gamma_\phi\,,\,\dot\Phi)$, \labelcref{eq:PairGamm-phiGauge} with the flowing field \labelcref{eq:FlowNielsenSolution} and the related target action, computed from the flow \labelcref{eq:FlowGqu}, is qualitatively different from the pair of the 'radical' classical target action flow. With \labelcref{eq:FlowNielsenSolution} the flowing field \textit{lifts} an approximation and re-instates the exactness of the flow. The dynamics is still largely stored in the flow itself. While this correction is an important one as the approximation is known to fail in the cases mentioned above, it is still a correction. Moreover, in \Cref{sec:OneLoopPIGs} we shall use the known failures of the background field approximation already at one-loop in order to show that the present approach rectifies these problems. 

This concludes the discussion of property \textbf{(i)}. It remains to show that there is a solution $\dot\Phi$ of \labelcref{eq:FlowNielsenSolution} with the properties \textbf{(ii,iii)}, that solves \labelcref{eq:FlowdotPhiProperty-iii} and \labelcref{eq:FlowNielsen}.

%%%%%%%%%%%%%%%%%%%%%%%%%%%
\subsection{Property \textbf{(ii)}: Gauge invariance}
\label{sec:Flowingii}

Background gauge invariance has been introduced as an auxiliary gauge invariance and the general solution of the flow of the trivial Nielsen identity, \labelcref{eq:FlowNielsenSolution}, also maintains background gauge invariance. To show this we use the flow of \labelcref{eq:Property-ii},  
\begin{align}
	\bar \delta_\omega\, \partial_t \Gamma_{\phi,\textrm{qu}}[\bar\varphi+\Phi]=0\,, 
	\label{eq:FlowProperty-ii}
\end{align}
where we have used that $\partial_t \Gamma_{\phi}= \partial_t \Gamma_{\phi,\textrm{qu}}$. The flow $ \partial_t \Gamma_{\phi,\textrm{qu}}[\bar\varphi+\Phi]$ with the property \textbf{(i)} is given by \labelcref{eq:FlowGqu} and all flows with 
\begin{align}
	\bar \delta_\omega\, F_\phi[\bar\varphi+\Phi]=0\,, 
	\label{eq:Constraint1Fphi}
\end{align}
satisfy \labelcref{eq:FlowProperty-ii}. This concludes our proof of background gauge invariance. However, the latter implies gauge invariance, \labelcref{eq:GaugeInvariantPIG}, as already discussed  in  \Cref{sec:FlowingGaugeInvariance}.

%%%%%%%%%%%%%%%%%%%%%%%%%%%
\subsection{Property \textbf{(iii)}: BRST symmetry}
\label{sec:Flowingiii}

With the properties \textbf{(i,ii)} we have shown that the current approach is background gauge invariant and satisfies the trivial Nielsen identity. It is left to show that the approach also has the property \textbf{(iii)} and hence its effective action satisfies the standard master equation for BRST invariance. This connects background gauge symmetry, and hence gauge symmetry, at each RG step to background gauge invariance, thus elevating the gauge symmetry of the target action from an auxiliary symmetry to the physical gauge symmetry. The flow of \labelcref{eq:Property-iii} is given by 
\begin{subequations} 
\label{eq:FlowProperty-iii}
\begin{align} 
		{\cal M}^{\ }_{\Gamma_\phi}[\bar\varphi,\Phi] \, \partial_t \Gamma_{\phi,\textrm{qu}} 
		=0 \,, 
		\label{eq:FlowProperty-iii1}
\end{align}
where once more we have used that $\partial_t \Gamma_{\phi}= \partial_t \Gamma_{\phi,\textrm{qu}}$. The linear operator ${\cal M}_{\Gamma_\phi}$ is given by 
\begin{align} 
	{\cal M}^{\ }_{\Gamma_\phi}=  \frac{\delta \Gamma_\phi}{\delta \mathcal{Q}_a}  \frac{\delta }{\delta \Phi_a} + 	\frac{\delta \Gamma_\phi}{\delta \Phi_a}  \frac{\delta }{\delta \mathcal{Q}_a} \,.
	\label{eq:calM}
\end{align} 
\end{subequations} 
The flow of the quantum part of the effective action, $\partial_t \Gamma_{\phi,\textrm{qu}}$, is given by the right-hand side of \labelcref{eq:GenGaugeFlowEquation}. Hence, \labelcref{eq:FlowProperty-iii} can be rewritten as a constraint equation for $\dot\Phi$, or rather for $F_\phi$. We find 
\begin{align} 
	{\cal M}^{\ }_{\Gamma_\phi} F_\phi[\bar\varphi+\Phi]= -	{\cal M}^{\ }_{\Gamma_\phi} \textrm{Flow}[\bar\varphi+\Phi]\,, 
	\label{eq:FlowdotPhiProperty-iii}
\end{align}
with the homogeneous solution $F_\phi$ of the flow of the Nielsen identity in \labelcref{eq:FlowNielsenSolution} and the flow term $\textrm{Flow}[\bar\varphi+\Phi]$ defined in \labelcref{eq:FlowExplicit}. \Cref{eq:FlowdotPhiProperty-iii} is a constraint equation for $F_\phi$ and while we have the trivial solution $F_\phi=\textrm{Flow}[\bar\varphi+\Phi]$ (classical target action), practical applications aim for the minimal solution. As a first evaluation we rewrite the right-hand side of \labelcref{eq:FlowdotPhiProperty-iii} by using that ${\cal M}^{\ }_{\Gamma_\phi} $ is a linear operator. We arrive at 
\begin{align} \nonumber 
 	{\cal M}^{\ }_{\Gamma_\phi} F_\phi[\bar\varphi+\Phi]=&  -\frac12 G_{ab}\Biggl[  {\cal M}^{\ }_{\Gamma_\phi} \partial_t R_k^{ab} \\[1ex] 
& 	\hspace{-1.5cm}- \left({\cal M}^{\ }_{\Gamma_\phi} \left[\Gamma_\phi^{(2)}+ R_k\right]^{bd}\right)\, \left( G\partial_t R_k\right)^{ad}\Biggr]    \,. 
	\label{eq:FlowdotPhiProperty-iiiMinimal}
\end{align}
\Cref{eq:FlowdotPhiProperty-iiiMinimal} vanishes at $c,\bar c,b=0$. Accordingly, the minimal solutions $F_\phi$ also vanish at $c,\bar c,b=0$ and only carry the BRST-rotations in the configuration space and no physical dynamics. This concludes our discussion of property \textbf{(iii)}.

%%%%%%%%%%%%%%%%%%%%%%%%%%%%
\subsection{Gauge invariant PIRGs in a nutshell}
\label{sec:PIGsinNutshell}

We close this Section with a collection of the equations from \Cref{sec:Flowingi,sec:Flowingii,sec:Flowingiii} required for a practical application of gauge invariant and gauge-consistent PIRGs.  

The final generalised flow equation for the quantum part of the effective action is given by 
\begin{subequations} 
\label{eq:PIG} 
\begin{align}\nonumber 
	\partial_t \Gamma_{\phi,\textrm{qu}}[\bar\varphi+\Phi] =&\, \frac12 G_{ab}[\bar\varphi+\Phi]\,\partial_t R_k^{ab}[\bar\varphi+\phi]\\[1ex] 
	&+F_\phi[\bar\varphi+\Phi]\,,
	\label{eq:PIGflow}
\end{align}
with the propagator $G[\bar\varphi+\Phi]$ in \labelcref{eq:Gvarphi+Phi}. \Cref{eq:PIGflow} satisfies the properties \textbf{(i)} and \textbf{(ii)} by construction, if $F_\phi$ is gauge invariant, \labelcref{eq:Constraint1Fphi}. Property \textbf{(iii)} requires a correction term $F_\phi[\bar\varphi+\Phi]$ with \labelcref{eq:FlowdotPhiProperty-iii}. 

The generalised flow for the effective action $\Gamma_\phi$ in \labelcref{eq:Property-i} is accompanied with the constraint equation \labelcref{eq:FlowNielsenSolution} for the flowing field. For the sake of completeness of the equations here we recall it
\begin{align} 
	\dot{\Phi}_a \frac{\delta \Gamma_\phi}{\delta  \Phi_a}  - G_{ab}\,\frac{\delta \dot{\Phi}_c }{\delta \Phi_b} \, R_k^{ac}
	= \Delta\textrm{Flow} -F_\phi[\bar\varphi+\Phi]\,, 
	\label{eq:FlowNielsenSolutionRecall}
\end{align}
\end{subequations} 
with $ \Delta\textrm{Flow}$ in \labelcref{eq:DefFlow}. The set of equations \labelcref{eq:PIGflow,eq:FlowNielsenSolutionRecall} and the flow of the master equation \labelcref{eq:FlowdotPhiProperty-iii} can be viewed as a set of a decoupled flow equations for the effective action and a constraint equation for $\dot\Phi$ that can be solved using an existing solution of the flow. This interpretation allows for a comprehensive discussion of the standard background field approximation: \\[-2ex]
 
The first term on the right-hand side is the flow in the background field approximation commonly used in quantum gravity. The term $F_\phi[\bar\varphi+\Phi]$ is a correction term. 
A radical choice would be $F_\phi=0$, for which the flow of $	\partial_t \Gamma_{\phi,\textrm{qu}}$ is precisely that of the background field approximation 
\begin{subequations} 
\label{eq:ExactBackground} 
\begin{align}
		\partial_t \Gamma_{\phi,\textrm{qu}}[\bar \varphi+ \Phi] =  \frac{1}{2} G_{ab}[\bar \varphi+\Phi]\,\partial_t  R_k^{ab}[\bar \varphi+\phi] \,,    
		\label{eq:ExactBackgroundFlow} 
\end{align}
with the flowing field 
\begin{align} 
		\dot{\Phi}_a \frac{\delta \Gamma_\phi}{\delta  \Phi_a}  - G_{ab}\,\frac{\delta \dot{\Phi}_c }{\delta \Phi_b} \, R_k^{ac}
		= \Delta\textrm{Flow} \,, 
		\label{eq:FlowNielsenSolutionBackground}
\end{align}
\end{subequations} 
and $ \Delta\textrm{Flow}$ in \labelcref{eq:DefFlow}. Note however that this choice violates property \textbf{(iii)} as this choice violates \labelcref{eq:FlowdotPhiProperty-iiiMinimal}. 
Still, the properties  \textbf{(i,ii)} persist and the respective flow is exact. In short, the present PIRG setup elevates the background field approximation, also including general proper times flows for all quantum field theories, not only gauge theories, to an exact flow equation. 
However, as we shall see in detail in the discussion of one-loop flows and beyond in \Cref{sec:OneLoopPIGs}, the flowing field then carries part of the physical dynamics and even part of the relevant terms. 
In the terminology put forward and discussed in \cite{Litim:2002xm}, the embedding in a PIRG clarifies that background flows and proper time RG flows are \textit{consistent} but not \textit{complete} flows: In complete flows, the complete (physical) dynamics is carried by the flow. In consistent flows only part of the dynamics is carried by the flow, but one has complete access to the full dynamics. As exact flows, PIRG flows are always consistent but they are only complete if the flowing field does not carry part of the dynamics. 

This general discussion already reveals that the choice $F_\phi=0$ does not result in a minimal flowing field $\dot \Phi$. On the contrary the flowing field potentially collects a sizeable part of the physical dynamics. In this case the flow $\dot \Phi$ or its integral $\int_t \dot \Phi$ can run into singularities. This already explains the qualitative differences observed between the fixed point properties in the background field approximation and those in the fluctuation approach, where this approximation is resolved. These differences are already present in pure quantum gravity (flow of the graviton mass parameter), but lead to qualitative differences in the matter-dependence of the fixed points, see the review \cite{Pawlowski:2020qer} and the literature cited there. A qualitative discussion of these issues will be done in \Cref{sec:OneLoopPIGs} at the example of one-loop flows. However, a full analysis of the flow $\dot\Phi$ in these cases will be presented elsewhere, and we simply note that a minimal solution for $\dot \Phi$ certainly avoids these difficulties. This concludes our discussion of the three properties underlying gauge invariant and gauge-consistent PIRGs.

%%%%%%%%%%%%%%%%%%%%%%%%%%%%
\section{One-loop gauge invariant PIRGs and beyond} 
\label{sec:OneLoopPIGs}

We elucidate the structure of gauge invariant flows with a detailed one-loop analysis. This analysis has direct consequences beyond one-loop level, and in particular also for non-perturbative applications including fixed point investigations. 

We shall first solve \labelcref{eq:PIG} explicitly in Yang-Mills theory for two choices of $F_\phi$. The first one is $F_\phi=0$, where the gauge invariant flow equation \labelcref{eq:PIG} reduces to the flow in the background field approximation \labelcref{eq:ExactBackground}. 
We note that \labelcref{eq:ExactBackground} may only be gauge-consistent in specific approximations and the right-hand side may not vanish in the general setup with higher order terms in the ghosts. 
Then, property \textbf{(iii)} requires flows with $F_\phi\neq 0$ that satisfy  \labelcref{eq:FlowdotPhiProperty-iiiMinimal}. At one-loop we can pull out the $t$-derivative, arriving at 
\begin{align} 
	{\cal M}_{\Gamma_\phi}F_\phi = - \frac12 \partial_t \left( G_{ab} \, {\cal M}^{\ }_{\Gamma_\phi} \left[\Gamma_\phi^{(2)}+ R_k\right]^{ab}\right)\,,
	\label{eq:MF}
\end{align} 
where we have used that $\partial_t   {\cal M}^{\ }_{\Gamma_\phi}$ is at least one-loop. Its insertion into another loop leads to two-loop terms and higher loops. Note that the $t$-derivative in \labelcref{eq:MF} has to be taken before performing the momentum or space-time trace. 

The second choice is more interesting for direct applications to non-perturbative phenomena including fixed points: We choose $F_\phi$ such that the ultraviolet or infrared relevant contributions in the flowing field $\dot\Phi$ are minimised. In \Cref{sec:OneLoopYM-PIG} we elucidate this at the explicit one-loop example. Here we only indicate the idea behind this minimisation. To that end we consider an expansion of the quantum part of the effective action in gauge invariant operators ${\cal O}_i$ with 
\begin{align}
	\Gamma_{\phi,\textrm{qu}}[\bar\varphi+\Phi] = \sum_{i\in \mathbbm{N}} {\cal C}^{(i)}_{a_1\cdots a_{n_i} }  {\cal O}_i^{a_1\cdots a_{n_i}}\,,
\label{eq:GquExpand}
\end{align}
with the coefficients ${\cal C}^{(i)}$. The flow of these coefficients provide the full dimensions of these operators $\gamma^{(k)}_{{\cal O}_i}$ and carry their relevance or irrelevance in the ultraviolet or the infrared. If the flowing field $\dot\Phi$ contributes to the anomalous dimension of UV or IR-relevant operators, one has stored a sizeable if not crucial part of the dynamics of the theory in the reparametrisation. Thus, structurally, we aim at a \textit{minimal} solution $\dot\Phi$ with the properties \textbf{(i,ii,iii)}, 
\begin{align} 
\left[	\dot{\Phi}_a \frac{\delta \Gamma_\phi}{\delta  \Phi_a}  - G_{ab}\,\frac{\delta \dot{\Phi}_c }{\delta \Phi_b} \, R_k^{ac}\right]_\textrm{rel} =0\,. 
\label{eq:IrrelevantdotPhi}
\end{align} 
The subscript ${}_\textrm{rel}$ indicates that \labelcref{eq:IrrelevantdotPhi} does not contain contributions to the relevant operators ${\cal O}_i^{\textrm{rel}}$, see also the discussion in \Cref{sec:PIRGProperties} and \Cref{fig:CompleteConsistentGphi}. As already mentioned above, we shall discuss this in detail in \Cref{sec:OneLoopYM-PIG} and \Cref{sec:OneLoopGravity-PIG}, including its practical implementation. If we can successfully set up such flows, it allows to use the non-perturbative results and fixed point properties of the gauge invariant PIRG directly, in general this requires the judicious supervision of the dynamics content of the flowing field.

%%%%%%%%%%%%%%%%%%%%%%%%%%%%
\subsection{Gauge invariant Yang-Mills theory} 
\label{sec:OneLoopYM-PIG}

We use one-loop perturbation theory in Yang-Mills theory for an illustration of the above points and in particular this analysis will elucidate \labelcref{eq:IrrelevantdotPhi}. We note in passing that reproducing one-loop perturbation theory is a common test of new functional setups in gauge theories. Needless to say, in a one-loop exact setup such as the fRG it is not a real test of \textit{consistency} or \textit{completeness} of an fRG as one-loop perturbation theory is hard-wired, for a detailed discussion see \cite{Litim:2001ky,Litim:2002xm, Litim:2002hj, Litim:2002ce, JMP2002}. However, as shown in \cite{Litim:2002ce, JMP2002}, one-loop universality is lost in the background field approximation if using infrared singular regulators, for a discussion in gravity see \cite{Folkerts:2011jz, Pawlowski:2020qer}. 

We proceed with the derivation of the one-loop $\beta$-functions from \labelcref{eq:PIG}, both for background field flows  \labelcref{eq:ExactBackground} with $F_\phi=0$ and for flows with a minimal $F_\phi$ with \labelcref{eq:IrrelevantdotPhi}. We shall see that in the former case one-loop universality is lost while in the latter case it is guaranteed. The flow equation for $\Gamma_{\phi,\textrm{qu}}$ for background field flows has first been done in \cite{Reuter:1993kw} and it can readily be performed with heat kernel techniques, see also \cite{Gies:2002af} for a full resummation relevant for non-perturbative applications in the infrared. We do not go into the details and quote the result for general regulators in \cite{Litim:2002ce} for axial gauges and \cite{JMP2002} for the presently used covariant gauges. For the present purpose, regulators can be classified by their infrared limit, 
\begin{align} 
	R_{A,c}(x\to 0) \propto \frac{k^2}{x^{n_{A,c}}}\,, \qquad x= - \frac{D_\mu^2}{k^2}\,, 
	\label{eq:IRlimits} 
\end{align}
for both ghost and gluon regulators. For $n_{A,c}>´ 0$ the regulators in \labelcref{eq:IRlimits} are infrared divergent. This only leads to an even more pronounced suppression of the infrared propagation and poses no problem in standard momentum cutoffs. However, in 	\labelcref{eq:PIGflow} the background field dependence of the propagator, $R_k=R_k[\bar\varphi]$, has been elevated to a dependence on the full field, $R_k[\bar\varphi]\to R_k[\bar\varphi+\phi]$. Evidently, this potentially changes the infrared dynamics encoded in the effective action $\Gamma_\phi$ in Yang-Mills theory. 

We initiate this analysis by recalling the standard argument that relates the one-loop $\beta$-functions of marginal couplings in $k$, $\beta^{(k)}_{g_i}$ to that in $\mu$, $\beta^{(k)}_{g_i}$. Here, $\mu$ is the RG-scale of the physical theory. More generally, this argument relates all one-loop anomalous dimensions $\gamma^{(k)}_{{\cal O}_i}$ of marginal operators ${\cal O}_i$ (including the $\beta$-functions) to the anomalous dimensions $\gamma^{(\mu)}_{{\cal O}_i}$ of the underlying physical theory at $k=0$. We find 
\begin{align} 
\gamma^{(k)}_{{\cal O}_i,0}= \gamma^{(\mu)}_{{\cal O}_i,0}\,,
\label{eq:gammamuk}
\end{align}
where the subscript ${}_0$ indicates the leading one-loop term. This identity does not hold any more at two-loop level, see e.g.~\cite{Pernici:1998tp, Pawlowski:2005xe, Rosten:2010vm}, and originates in the absence of further scales. As already mentioned before, it has been shown in \cite{Litim:2002ce,JMP2002} that this identity does not hold true for the $\beta$-function $\beta_{\alpha_s}$ of the strong coupling $\alpha_s=g_s^2/(4 \pi)$ in Yang-Mills theory for flows in the background field approximation \labelcref{eq:ExactBackground} with singular infrared regulators. Moreover, in this context it is important to distinguish between the $\bar\varphi$-dependence triggered by the gauge fixing part of the theory, $S_\textrm{gauge}$, and that from the regulator. This is deferred to \Cref{app:StandardBack1loop}, where we briefly review these differences for the standard background fRG approach with $\Gamma_a[A,\Phi_f]$ and $\phi_\mu=a_\mu$. Specifically we discuss the origin and content of the different terms in the modified Nielsen identity. This elucidates the mechanisms behind the present gauge invariant PIRG and the rôle and physics content of the flowing field.  

We proceed with the explicit computation. To begin with, in a gauge invariant setup with the fundamental fluctuation field $\phi=a$ with $\dot\phi=0$, we expect 
\begin{align} 
	\Gamma_{a,\textrm{qu}}^{\textrm{1-loop} }[A]= \frac{Z_F}{2} \int_x \tr F_{\mu\nu}^2[A] + \cdots \,,
\end{align} 
where $\cdots$ stands for higher order gauge invariant terms in the field $A=\bar A+a$ in the expansion \labelcref{eq:GquExpand}. The $\beta$-function $\beta_{\alpha_s}^{(k)}$  follows from gauge invariance: $\partial_t ( Z_F Z_g^2 )=0$, where $Z_g$ is the dressing of the gauge coupling. This leads to 
\begin{align}
\beta^{(k)}_{\alpha_s}= - k\frac{d \log Z_F}{d k}  \,. 
\label{eq:kbeta}
\end{align} 
At one-loop and in the absence of further scales we use  $\beta^{(k)}_{\alpha_s}=\beta^{(\mu)}_{\alpha_s}$ and the latter $\beta$-function is given by 
\begin{align} 
	\beta^{(\mu)}_{\alpha_s} = \mu\frac{d \log \alpha_s}{d\mu}\,, \qquad  \beta_0= \beta^{(\mu)}_0=- \frac{22}{3} N_c \,   \frac{\alpha_s }{4 \pi }\,. 
	\label{eq:beta1loop} 
\end{align}
An explicit computation with the background PIRGs \labelcref{eq:ExactBackground} leads us to 
\begin{align} 
	\partial_t  \Gamma^{\textrm{1-loop}}_\phi  =& - \beta^{(k)}_{0}  \frac{Z_F}{2} \int_x \tr F_{\mu\nu}^2 +\cdots \,,
	\label{eq:BackFlowOneLoop} 
\end{align}
where $\cdots$ stands for the higher order terms in \labelcref{eq:GquExpand}. In \labelcref{eq:BackFlowOneLoop} we have already anticipated the fact that 
the  flow \labelcref{eq:ExactBackground} cannot hold true for the effective action of the fundamental field. Indeed, the one-loop $\beta$-function with respect to $k$ reads 
\begin{align} 
 \beta^{(k)}_0 	=   \left( 1 +  \frac{10 n_A + n_c }{11} \right)\, \beta_0\,. 
\label{eq:beta0k}
\end{align}
This result was obtained in \cite{Litim:2002ce,JMP2002} and signals the incompleteness of background field flows. 
According to the classification into complete or consistent RG flows \cite{Litim:2002xm} discussed in \Cref{sec:PIRGProperties}, flows with $n_{A,c}\neq 0$ are only consistent flows in the PIRG approach, as part of the dynamics of the relevant operator $\tr F^2$ is encoded in the flowing field $\dot\phi\neq 0$. 
Hence the above results cannot be an exact result for $\Gamma_a$: otherwise the one-loop $\beta$-function would be the universal one for all regulators, and we conclude $\phi\neq a$. 
Note that this failure is less obvious for mass-type regulators with $n_A=n_c=0$. However, as the derivation of \labelcref{eq:BackFlowOneLoop} shows, it cannot hold true for the effective action of the fundamental fluctuation field. 

Evidently, the cutoff $\beta$-function \labelcref{eq:beta0k} only reduces to the universal one-loop one for mass-like regulators with $n_A=n_c=0$. 
As discussed in \cite{Litim:2002ce, JMP2002}, this originates in the existence of another implicit scale for IR-singular regulators. 
While this is seemingly artificial for Yang-Mills theory at one loop, it illustrates the potential generic failure of simply elevating the background field in the regulator to a full dynamical field. 

In the present approach this difference is contained in the flowing field that solves the Nielsen identity \labelcref{eq:FlowNielsenSolution}. 
This entails that part of the flow carries a non-trivial dependence on the cutoff even for the 'universal' terms. Indeed, in \cite{Litim:2002ce, JMP2002} it has been shown that 
\begin{align} \nonumber 
&	\left[\left(\frac{\delta }{\delta a}- \frac{\delta }{\delta \bar A}\right) \Bigl( \Delta\textrm{Flow} \Bigr)\right]_{a=0}\\[2ex]
&\hspace{2.5cm}=   \beta_0 \frac{10 n_A + n_c }{11} \,\frac{\delta S_A[A]}{\delta A}+\cdots\,. 
\label{eq:OneLoopCorrection}
\end{align}
Consequently the flowing field term in the flow carries terms that are proportional to the (only) UV-relevant or marginal term in Yang-Mills theory, 
\begin{align} \nonumber 
&\left(\frac{\delta }{\delta a}- \frac{\delta }{\delta \bar A}\right) \left[	\dot{\phi}_a \frac{\delta \Gamma_\phi}{\delta  \phi_a}  - G_{ab}\,\frac{\delta \dot{\phi}_c }{\delta \phi_b} \, R_k^{ac}\right]_{a=0} \\[2ex] 
&	\hspace{2.5cm}=  \beta_0 \frac{10 n_A + n_c }{11} \,\frac{\delta S_A[A]}{\delta A}+\cdots\,. 
	\label{eq:FlowNielsenSolutionYM}
\end{align}
In conclusion, in the background field PIRG we have stored part of the relevant running of the $\beta$-function in the field transformation. 
While this is certainly possible, it complicates the assessment of the physics dynamics, and in particular the search for fixed points and their stability. 
Moreover, it also makes the direct access to physics intricate. 
Note also that this can be used to even change the sign of $\beta$-functions or rather their part included in the effective action. 

The embedding in the exact PIRG setup rectifies these deficiencies and leaves us with two clear recipes for using background field approximation flows:
\begin{enumerate} 
\item The first option is to introduce a flowing field which arranges for the background field approximation, that is $F_\phi=0$. Then one controls the flow with the constraint equation for $\dot\Phi$. If the latter runs into a singularity or gets sizeable, the results of the background field approximation have to be interpreted with care. Moreover, one can restore the standard flow by resolving the Nielsen identity encoded in $\dot \Phi$. 
\item The second option is to use the \textit{minimal} solution \labelcref{eq:IrrelevantdotPhi}. To approach it practically, it is suggestive to minimise the RG-running of the flowing field $\dot\phi$. In short, we demand \labelcref{eq:IrrelevantdotPhi}. More concretely, we want to remove the violations of \labelcref{eq:IrrelevantdotPhi} triggered by the $\phi$-dependence of the regulator in the PIRG flow, while we keep the completion property related to the gauge part of the action discussed in \Cref{app:StandardBack1loop}. This leads to the constraint on $F_\phi$ with 
\begin{align} 
\left[ \dot{\Phi}_a \frac{\delta \Gamma_\phi}{\delta  \Phi_a}  - G_{ab}\,\frac{\delta \dot{\Phi}_c }{\delta \Phi_b} \, R_k^{ac}\right]_{(\tr F^2)_{R_k}^{\ }} \stackrel{!}{=}0\,. 
\label{eq:ZeroRelevant}
\end{align} 
The subscript ${}_{(\tr F^2)_{R_k}^{\ }}$ in \labelcref{eq:ZeroRelevant} indicates a procedure which removes the contributions to $\int \tr F^2$ that are sourced in the $\phi$-dependence of the regulator. We only eliminate this term as it is the only relevant gauge invariant operator in Yang-Mills theory.  
\end{enumerate} 
Now we implement the second option at one-loop. Then, \labelcref{eq:ZeroRelevant} reduces to 
\begin{align} 
\left. \dot{\Phi}_a \frac{\delta \Gamma_\phi}{\delta  \Phi_a}  \right|_{({\tr F^2})^{\ }_{R_k}} \stackrel{!}{=}0\,.
	\label{eq:IrrelevantdotPhiYM}
\end{align} 
This term has been already discussed in detail in \cite{Litim:2002ce,JMP2002}: It is encoded in  
\begin{align} 
	\frac{\delta R^{bc}_k[\bar\varphi+\phi]}{\delta\bar \varphi_a} \frac{\delta \Gamma_\phi }{\delta R^{bc}_k} = \frac12 G_{bc} \frac{\delta R^{bc}_k[\bar\varphi+\phi]}{\delta\bar\varphi_a}  \,, 
	\label{NielsenRk}
\end{align}
where we dropped a term proportional to the $R_k$ variation of $\hat\phi_k$. 
Any subtleties related to the latter term are avoided by simply taking the respective derivative of \labelcref{eq:FlowNielsenSolutionBackground}. 
This leads us with a practical relation for a minimal $F_\phi$ in Yang-Mills theory,  
\begin{align} 
	\frac{\delta R^{bc}_k}{\delta\bar\varphi^a} \frac{\delta }{\delta R^{bc}_k}\left[	\dot{\Phi}_a \frac{\delta \Gamma_\phi}{\delta  \Phi_a}  - G_{ab}\,\frac{\delta \dot{\Phi}_c }{\delta \Phi_b} \, R_k^{ac}\right]_{(\tr F^2)_{R_k}^{\ }}=0 \,. 
	\label{eq:FlowNielsenSolutionBackgroundRk}
\end{align}
\Cref{eq:FlowNielsenSolutionBackgroundRk} implies that 
\begin{align} 
	\frac{\delta R^{bc}_k}{\delta\bar\varphi^a} \frac{\delta }{\delta R^{bc}_k}\Bigl[	\Delta \textrm{Flow}[\bar\varphi,\Phi] - F_\phi[\bar\varphi+\Phi]\Bigr]_{(\tr F^2)_{R_k}^{\ }}\!\!\!=0\,.
\end{align}
Now we evaluate \labelcref{eq:FlowNielsenSolutionBackgroundRk} at one-loop and use that we can safely evaluate $F_\phi$ at $\Phi=0$. Then we arrive at 
\begin{align} 
F_\phi[\bar\varphi+\phi] =- \frac12 f^{aa}_\phi(x)\,, 
\label{eq:FphiUniversal}
\end{align} 
where $x$ is the Laplacian of the gauge field and ghost. \Cref{eq:FphiUniversal} only depends on the gauge field and not on the auxiliary fields. \Cref{eq:FphiUniversal} is the trace of the operator $f^{ab}_\phi$, and at one-loop we find  
\begin{align} 
 \frac{\partial f^{ab}_\phi(x)}{\partial x} = \left. \partial_t \left( G_{ad}(x) \frac{\partial R_k(x)^{bd}}{\partial x}\right)\right|_{c,\bar c, b=0} \,. 
\label{eq:DefUniversalf}
\end{align} 
In \cite{Litim:2002ce,JMP2002} this term was extracted as a deviation of the background field approximation from an exact flow, in the PIRG approach it is a term that can be accommodated with the flowing field. 
We emphasise that for regulators with $n_{A,c}\neq 0$ in \labelcref{eq:IRlimits}, the $t$-derivative has to be applied first before performing the trace by summing over space-time, hence the definition of $f^{ab}_\phi$.  
If the constraint \labelcref{eq:FphiUniversal} is to hold beyond one-loop, \labelcref{eq:FphiUniversal} has to be improved by resummation terms. 
This,  further improvements and non-perturbative applications will be discussed in \cite{IKMP}. 

\Cref{eq:DefUniversalf} is gauge invariant. Hence, it is trivially BRST invariant as it only depends on the gauge field, and the BRST transformation of the gauge field is a gauge transformation. Beyond one-loop, the full gauge invariant flow is given by 
\begin{align}\nonumber 
	\partial_t \Gamma_{\phi,\textrm{qu}}[\bar \varphi+ \Phi] =  &\, \frac{1}{2} G_{ab}[\bar \varphi+\Phi]\,\partial_t  R_k^{ab}[\bar \varphi+\phi] \\[1ex] 
	& +F_\phi[\bar \varphi+\Phi]\,. 
	\label{eq:ExactGauge} 
\end{align}
The functional $F_\phi$ removes the relevant or rather marginal part in $\dot\Phi$, see \labelcref{eq:IrrelevantdotPhiYM}. Then, the one-loop $\beta$-function reduces to the known universal result with 
\begin{align}
	\beta^{(k)}_{0} = \beta_0\,,
\end{align}
with the universal $\beta_0$ in \labelcref{eq:beta1loop}. 
This leaves us with a gauge invariant PIRG in non-Abelian gauge theories, where the UV-relevant dynamics is governed by the flow and not by the parameterisation. 
We will discuss the non-trivial infrared sector with its non-perturbative confining physics elsewhere \cite{IKMP}.

%%%%%%%%%%%%%%%%%%%%%%%%%%%%
\subsection{Diffeomorphism invariant gravity} 
\label{sec:OneLoopGravity-PIG}

In this Section we discuss the application of the gauge invariant PIRG to quantum gravity, for a recent review see the asymptotic safety chapter in \cite{Bambi2024-nm}. 
The derivations and results in Yang-Mills theory in \Cref{sec:OneLoopGravity-PIG} already teach some lessons for quantum gravity. 
First of all, the crucial relevance of \labelcref{eq:IrrelevantdotPhi} is imminent in quantum gravity: a conclusive analysis of the Reuter fixed point in metric gravity relies on the control of the UV-relevant sector of the theory. 
It is suggestive that the observed qualitative differences in the flow equation of the cosmological constant in the background field approximation and that for the mass parameter of the graviton in the fluctuation approach are sourced in relevant contributions to the flowing field $\dot\phi_{\mu\nu}$. 
The same reasoning applies to the singularity of the background flow equations for $\lambda=1/2$, where $\lambda=\Lambda/k^2$ is the cosmological constant $\Lambda$ in \labelcref{eq:EH-Action} in \Cref{app:BackgroundYM}, measured in units of the the cutoff scale. This singularity is absent in the fluctuation approach with $\dot\phi_{\mu\nu}=0$. 

The above analysis suggests that for $F_\phi=0$ we should monitor the part of the physics dynamics that is stored (implicitly) in the flowing field $\dot\Phi$. 
This amounts to simply using the results for the background field flows in the literature and solving the differential equation for $\dot\Phi$ in \labelcref{eq:FlowNielsenSolutionBackground} on the solution of the background flow \labelcref{eq:ExactBackgroundFlow}. 
It is already clear from the results in the literature for the background field flows and those in the fluctuation approach that \labelcref{eq:FlowNielsenSolutionBackground} carries part of the flow of the cosmological constant. Hence, it is suggestive to use a non-vanishing $F_\phi$ that removes this part of the flowing field. This  keeps the full dynamics of the cosmological constant or the mass parameter of the graviton in the flow itself. 

This brings us to the second intricacy of background field flows, their singularity at $\lambda=1/2$. This singularity is caused by 
\begin{align} 
	\lim_{\lambda\to \frac12} G[\bar\varphi+\Phi] \to \infty \,,
	\label{eq:GBackInfty}  
\end{align}
for the propagator $G[\bar\varphi+\Phi] $ defined in \labelcref{eq:Gvarphi+Phi}. Indeed, this singularity is also caused by the fact that part of the flow of the mass parameter of the graviton is stored in 	\labelcref{eq:FlowNielsenSolutionBackground}. Consequently, the reparameterisation $\dot \phi_{\mu\nu}$ itself is singular at $\lambda=1/2$ as it is linearly depending on \labelcref{eq:GBackInfty} and \labelcref{eq:ExactBackgroundFlow} ceases to be an exact flow: it is neither complete nor consistent. 

These relations also entail an important conclusion that can be drawn from the present approach. Any potential physics content of this singularity such as the absence of a positive infrared value of the cosmological constant is not supported by the fRG approach to gravity: This readily follows from the following argument. All gravity flows in the literature can be interpreted as PIRG flows with a constraint for the flowing field $\dot\phi_{\mu\nu}$. Physics properties are only supported by the flows if they are accompanied with flowing fields. In the present case all flows showing this singularity have trivially singular flowing fields. We hasten to add that this does not entail that the singularity is unphysical but this interpretation is simply not supported by the current analyses. 

The important conclusion of this analysis is that background fields flows in metric gravity should be augmented with a non-trivial $F_\phi$ to allow conclusive statements about the physics of the Reuter fixed point. This applies in particular to matter-gravity systems as already assessed in \cite{Meibohm:2015twa,Pawlowski:2020qer}. Moreover, an appropriately chosen $F_\phi$ also eliminates the $\lambda=1/2$ singularity in the flow. A trivial solution is to enforce the fluctuation approach flow of the graviton mass parameter for the cosmological constant in the gauge invariant PIRG. Evidently this eliminates a specific relevant part of the flowing field $\dot\phi_{\mu\nu}$ including the removal of the parametrisation singularity. However, such a brute force solution makes the physics interpretation or rather the reconstruction \cite{Ihssen:2024ihp} of the cosmological constant in the PIRG setup more intricate. Instead it is suggestive to use a variant of the $F_\phi$ \labelcref{eq:DefUniversalf}, that follows from the relevance constraint \labelcref{eq:IrrelevantdotPhi} in Yang-Mills theory. The modification of \labelcref{eq:DefUniversalf} follows from 
\begin{align} 
	\left[ \dot{\Phi}_a \frac{\delta \Gamma_\phi}{\delta  \Phi_a}  - G_{ab}\,\frac{\delta \dot{\Phi}_c }{\delta \Phi_b} \, R_k^{ac}\right]_{({\cal O}_i)_{R_k}} \stackrel{!}{=}0\,,
	\label{eq:IrrelevantdotPhiGravity}
\end{align} 
with $i=1,2,3$ for the three potentially relevant operators 
\begin{align} 
	{\cal O}_1  =\sqrt{g}\,,\qquad {\cal O}_2 =\sqrt{g}\, R\,,\qquad {\cal O}_3=\sqrt{g} \, R^2\,.
	\label{eq:RelevantOGravity}
\end{align} 
The respective couplings are the cosmological constant $\Lambda$ for $i=1$, the Newton constant for $i=2$ and the $R^2$ coupling $g^{\ }_{R^2}$ for $i=3$. Note that we can add further operators to this list. For a discussion of the relevance ordering in the fluctuation approach see \cite{Pawlowski:2020qer}, the application of this setup to quantum gravity with \labelcref{eq:IrrelevantdotPhiGravity} will be discussed in \cite{IKMPS}.

%%%%%%%%%%%%%%%%%%%%%%%%%%
\section{The flowing landscape of gauge theories} 
\label{sec:GaugeInvariantLandscape}

In this final Section we discuss the landscape of RG flows in gauge theories through the lens of the novel  PIRG approach to gauge theories. A recent comprehensive survey of fRG flows in gauge theories, including further literature, can be found in \cite{Dupuis:2020fhh}.

Functional renormalisation group (fRG) flow approaches in gauge theories have to satisfy simultaneously the requirement of implementing a mode cutoff (typically momentum modes) and that of gauge invariance. Moreover, if formulated for the 1PI effective action, the kinetic operator of the gauge field has to be inverted, which requires either a gauge fixing or a covariant and hence field-dependent projection on the physical degrees of freedom. This leaves us with the choice of a gauge-fixed setting or the formulation in gauge-covariant or gauge invariant degrees of freedom. 

The PIRG approach encompasses all different approaches with or without approximations, simply by defining the flow in a given approach as the target action flow. This also allows us to relate gauge-fixed and gauge invariant approaches. The brief discussion below is meant to provide a structural overview. It is far beyond the scope of the present work to provide a comprehensive embedding. Moreover, in the following we mostly concentrate on earlier suggestions for gauge invariant flows, but also discuss the link to standard gauge-fixed flows.

%%%%%%%%%%%%%%%%%%%%%%%%
\subsection{Gauge-fixed flows and modified master equation} 
	%Slavnov-Taylor identities} 
\label{sec:GaugeFixed+mSTIs}

The commonly used fRG approach to gauge theories is that with a gauge fixing on the level of the fundamental fields. Its flow is accompanied by the modified master equation \labelcref{eq:mMaster} for non-vanishing cutoff, see \Cref{app:BackgroundYM}. For the benefit of the reader we recall \labelcref{eq:mMaster} here, 
\begin{align}
	\frac{\delta \Gamma_k}{\delta \mathcal{Q}_i}  \frac{\delta \Gamma_k}{\delta \Phi_i}
	=
	R^{ij} G_{jl}  \frac{\delta^2 \Gamma_k[\Phi,\mathcal{Q}]}{\delta\Phi_l \delta Q^i}
	\, .
	\label{eq:mMasterMain}
\end{align}
This is the standard approach for quantitative studies in QCD, where it is used with the Landau gauge $\xi=0$, for recent reviews see \cite{Dupuis:2020fhh, Fu:2022gou}. In quantum gravity it is used in the fluctuation approach, for a recent review see \cite{Pawlowski:2020qer}. 

In short, in this approach one monitors the breaking of BRST invariance during the flow, which is encoded in the right-hand side of \labelcref{eq:mMasterMain}. 
Then, the effective action flows into the physical one at $k=0$, which satisfies the master equation without the breaking term, see e.g.~\cite{Pawlowski:2022oyq}. In practical applications one may simply tune the initial conditions of a flow such that the physical effective action at $k=0$ satisfies the master equation \labelcref{eq:mMasterMain} with a vanishing right-hand side. 
It is the computationally most developed approach and produces quantitative results for Yang-Mills theory and QCD, also at finite temperature and density. The BRST-breaking term on the right-hand side of \labelcref{eq:mMasterMain} triggers a (longitudinal) gluonic mass term $m_{A,k}^2\ \int A_\mu^2$, which is vanishing at $k=0$. Its implementation comes at the price of a quadratic fine-tuning. While this fine-tuning can be incorporated, it is computationally demanding. Moreover, the ongoing quest for quantitative precision required for even qualitative statements in high density and competing order regimes is computationally very demanding, and the quadratic fine-tuning amplifies these costs largely. 

Therefore it is very desirable to embed the standard gauge-fixed approach into a PIRG approach with the property \textbf{(iii)}, \labelcref{eq:Property-iii}. This is only a minimal variation of the standard approach but comes with a standard master equation \labelcref{eq:MasterBRST}. Then, the right-hand side of \labelcref{eq:mMasterMain} is vanishing and no mass term is present: this allows to use the sophisticated code framework for Landau gauge QCD based on \cite{ Ihssen:2023qaq, Sattler:2024ozv, Fu:2025hcm} without any change, but eliminates the quadratic fine-tuning problem. Moreover, it simplifies the construction of consistent approximations of the effective action in terms of BRST-invariants.   

We remark that it has been discussed in the literature, how the standard approach can be made gauge-consistent. A vanishing right-hand side in the modified master equation \labelcref{eq:mMasterMain} can be obtained by using composite operators and specifically the full two-point functions \cite{Pawlowski:2005xe}, see also \cite{Lavrov:2012xz}. For a systematic approach to the master equation and its fate in fRG approaches see \cite{Igarashi:2009tj}, for more recent works addressing the relation between fRG approaches for the 1PI effective action and the Wilson effective action see \cite{Igarashi:2016gcf, Igarashi:2019gkm}. 

The analysis in \cite{Pawlowski:2005xe} can also be interpreted or rewritten as absorbing the right-hand side in a composite field $\phi$. The present PIRG approach to gauge theories offers a practical implementation of this idea with a minimally flowing field. While one can embed the standard approach in a fully gauge invariant PIRG setup as introduced here, in a first step one may only consider property \textbf{(iii)} as discussed in \Cref{sec:Flowingiii}. 

We close this Section with a brief discussion of the fate of the mass term: Evidently, it is absent if the right-hand side of \labelcref{eq:mMasterMain} vanishes. We restrict ourselves to $k\to \infty$, where the effective action of Yang-Mills theory approximately reduces to the classical one, \labelcref{eq:Sclassical}, with $k$-dependent coefficients. Within a formulation with the fundamental fluctuations fields $\Phi_f$, the modified master equation \labelcref{eq:mMasterMain} entails that all couplings only agree up to minor modifications and we have a quadratically running mass term,   
\begin{align} 
\Gamma_{A,k\to \infty}[\Phi_f] \approx  S_\textrm{YM} [\Phi_f] + \frac 12 m_{A,k}^2 \int_x (A^a_\mu)^2 \,, 
\label{eq:FateofMass1} 
\end{align}
with the superfield of the fundamental fields $\Phi_f=(A,c,\bar c)$ in \labelcref{eq:Superfieldf}. For more details see \cite{Dupuis:2020fhh, Fu:2022gou} and references therein. In contrast, flowing fields $\dot\Phi$ with the property \textbf{(iii)} (standard BRST invariance) lead to 
\begin{align} 
	\Gamma_{\phi,k\to \infty}[\Phi] \approx  S_\textrm{YM} [\Phi]\,.  
	\label{eq:FateofMass2} 
\end{align}
Importantly, in this case the composite field $\Phi$ does not reduce to the classical one in the limit $k\to \infty$. Here it pays off that the gauge fixing in the PIRG approach to gauge theories has been defined on the level of the effective action and not on the level of the classical action in the path integral. 

The standard flow, or rather the flow term in the generalised flow equation, re-generates a mass term, since it has a contribution $\partial_t m_{A,k}^2\neq 0$, even if $m_{A,k}^2=0$ at this scale. In PIRG flows with the property \textbf{(iii)} this contribution is absorbed into the flowing field at each RG step. This entails a non-linear transformation of the flowing field with operators $(\partial_t m_{A,k}^2)\, G_{A,k}(p)$, with the gluon propagator $G_{A,k}(p)$. This transformation may be related to a non-linear massive gauge fixing \cite{Gies:2022mar}. In the presence of the infrared regularisation with $R_k$, this operator carries no infrared non-locality and vanishes for $k\to 0$. In short, such a formulation avoids the common non-localities of gauge-consistent regularisations. More details will be provided elsewhere.

%%%%%%%%%%%%%%%%%%%%%%%%
\subsection{Background field and proper time flows} 
\label{sec:Background+PTRG}

We keep this Section very brief, as we have mainly used exact background field flows for the illustration of the novel PIRG approach to gauge theories. Indeed, the first applications to gauge theories with the fRG approach were mostly done within the background field approach, and most of them resorted to the background field approximation \labelcref{eq:ExactBackground}. In the background field approximation the regulator depends on the full field $\bar\varphi+\phi$, and a subclass of these flows are the proper time flows for which the flows only depend on the operator $\Gamma^{(2)}[\bar\varphi+\Phi]$. See \cite{Litim:2002hj} for the representation of proper time RG flows in exact background field flows. There are two main application areas, low energy QCD and quantum gravity, and the respective founding papers are \cite{Reuter:1993kw} (QCD) and \cite{Reuter:1996cp} (gravity). To date, these flows has been mostly used as an approximation of an exact flow, see \cite{Litim:2002hj, Bonanno:2019ukb, Falls:2020tmj} for respective discussions. More recently, the relation to exact flows and the consequences of the Nielsen identities (aka split Ward identities) have been studied and exploited in e.g.~\cite{Dietz:2015owa, Labus:2016lkh, Mandric:2022dte, Pagani:2024lcn}. 

In the present work we have shown that background field flows are exact: they are consistent but incomplete flows according to the classification in \cite{Litim:2002xm} and the generalisation introduced here, see \Cref{sec:PIRGProperties} and \Cref{fig:CompleteConsistentGphi}. For more details we refer to the discussion in \Cref{sec:PIGsinNutshell} and \Cref{sec:OneLoopPIGs}. This embedding allows for an interpretation of the plethora of results with these flows in terms of a well-controlled approximation of exact flows, and the approximation is controlled by the flowing field $\dot\Phi$ in \labelcref{eq:ExactBackground}.  

Recently, a simplified flow equation has been put forward, \cite{Wetterich:2024ivi}, which involves a map from the Wetterich equation to proper time flows with flowing fields. It is also connected to the proposal \cite{Wetterich:2016ewc, Wetterich:2017aoy} aiming at the same task. The resulting equation in \cite{Wetterich:2024ivi} has the same form as the background field flow \labelcref{eq:ExactBackgroundFlow}, the only difference being the propagator \labelcref{eq:Gvarphi+Phi}, which is modified by $\Gamma^{(2)}_k \to \Gamma^{(2)}_k+{\cal Q}$. The defining equation of ${\cal Q}$ resembles the terms discussed in \cite{Litim:2002ce, JMP2002, Litim:2002hj}. In contradistinction, the differences of the flows in \cite{Wetterich:2016ewc, Wetterich:2017aoy} are additive and the ghost contribution is one loop exact. 

The contribution ${\cal Q}$ to the two-point function of the gauge field in \cite{Wetterich:2024ivi} as well as the additive flow terms in \cite{Wetterich:2016ewc, Wetterich:2017aoy} can be accommodated with specific $F_\phi$'s with the respective $\dot\Phi$ in the general PIRG for gauge theories \labelcref{eq:PIG}. Notably, the PIRG approach allows for all variations, ranging from a complete separation of the flow of the effective action and the constraint equation of the flowing field to a complete entanglement. Due to this generality all different suggestions are encompassed by it, and we consider this flexibility as its most attractive feature.

%%%%%%%%%%%%%%%%%%%%%%%%
\subsection{Gauge invariant flows and composites} 
\label{sec:GaugeInvariant+Composites}

In this final Section we discuss gauge invariant flows, most of which are also part of the survey \cite{Dupuis:2020fhh}. Roughly speaking, one may distinguish two different approaches, even though there are close connections between them. 

The first approach encompasses flows, based on gauge covariant variables akin to link variables in lattice gauge theories. This requires a covariant and hence field-dependent cutoff. These cutoffs can be introduced without further problems for general flows with field-dependent coarse-graining kernels for the Wilson effective action, the Wegner flow \cite{Wegner_1974}. For a field-independent coarse-graining kernel, the Wegner equation reduces to the Polchinski equation \cite{Polchinski1984}. The Wegner flow has been utilised in \cite{Morris:1999px, Morris:2000fs, Morris:2005tv} for the Wilsonian effective action of gauge theories, see also the review \cite{Rosten:2010vm}. More recently the lattice analogy was used more directly in \cite{Sonoda:2020vut, Miyakawa:2021wus, Sonoda:2022fmk, Miyakawa:2023yob, Haruna:2023spq}. In these works the fRG flow is formulated akin to the gradient flow on the lattice, utilising the diffusion form of the Wegner equation. 

The key idea of the PIRG approach is the formulation of flows in terms of the pair of the target action and flowing field \cite{Ihssen:2024ihp}. Accordingly, it can be formulated for both Wegner flows and the generalised flow equation \cite{Pawlowski:2005xe}, which is the Legendre transform of the Wegner equation, see \cite{Baldazzi:2021ydj, Ihssen:2022xjv}. Hence, its gauge theory version \labelcref{eq:GenGaugeFlows} also accommodates the flows in \cite{Morris:1999px, Morris:2000fs, Morris:2005tv,Sonoda:2020vut, Miyakawa:2021wus, Sonoda:2022fmk, Miyakawa:2023yob, Haruna:2023spq}. Indeed, the formulation of the latter gradient flow in terms of the 1PI effective action is specifically interesting also for machine learning applications as advocated in \cite{Cotler:2022fze}. 

The second approach encompasses flows that are based on gauge invariant degrees of freedom. These flows are trivially gauge invariant. One variant is based on the Vilkovisky-DeWitt action \cite{Vilkovisky:1984st, DeWitt:2003pm}, and has been put forward in \cite{Branchina:2003ek, Pawlowski:2003sk}. In this approach one singles out a section in the fibre bundle of the gauge theory and uses the transversal projection of the respective reparametrisation-invariant field $\sigma^a$ (section field) as the argument of the effective action. 
This construction is accompanied by the appropriate path integral measure. Moreover, since the section field transforms as a scalar under gauge transformations, one can readily couple it to an infrared cutoff. Then, gauge and reparametrisation invariance can be proven directly from the flow itself \cite{Pawlowski:2003sk}, analogously to the proofs of the properties \textbf{(i,ii,iii)} in \Cref{sec:Flowingi-ii-iii}. 
The use of a specific section comes with inherent non-localities that originate in the projection of a given field on the section. 
This is most obvious from the form of the generator of gauge transformations along the fibre, related to the ghost propagator. 
While the projection procedure is covariant, this is tantamount to a (non-linear) gauge fixing. This may even deform the one-loop running of the theory, see e.g.~\cite{DeWitt:1995cx,DeWitt:1998eq}. The respective non-localities are also present in dressed approaches with composite fields, see e.g.~\cite{Capri:2015ixa} and the recent work \cite{Falls:2025tid}. 
Naturally, these non-localities are under control within an fRG approach, for a discussion in the Vilkovisky-DeWitt approach see  \cite{Pawlowski:2003sk, Donkin:2012ud}. 
Moreover, the PIRG approach can be used for a flowing construction of the section fields $\sigma^a$ from the underlying gauge fields. 
Similar flowing fields can be constructed for dressed fields or gauge invariant degrees of freedom \cite{Frohlich:1981yi}, for recent developments see \cite{Egger:2017tkd, Maas:2017xzh, Sondenheimer:2019idq, Maas:2020kda}. 
This is an interesting option for a comprehensive investigation of Beyond the Standard Model theories without relying on the perturbative identification of gauge invariant resonances.

%%%%%%%%%%%%%%%%%%%%%%%%
\subsection{The practitioner's point of view} 
\label{sec:GaugeInvariantWrap}

We close \Cref{sec:GaugeInvariantLandscape} with a short note from a complementary practitioner's point of view. To begin with, we have argued that the PIRG approach encompassed both gauge-fixed and gauge invariant flows in the literature. 
This begs the natural question whether there is an optimal or preferred formulation. 
An equally natural and diplomatic answer would be 'it depends on the application'. 
Be this as it may, we would like to emphasise the following point already discussed in \cite{Dupuis:2020fhh}: gauge fixing is merely the choice of a parametrisation. Importantly, 
a covariant gauge fixing leads to \textit{local} and \textit{momentum-local} correlation functions, for a discussion see \cite{Pawlowski:2020qer, Ihssen:2024miv}. If at all, an optimal choice of a flow for a given application is directly related to an optimised parametrisation. 

The task of finding an optimal parametrisation has both technical and conceptual aspects. If striving for results with a small systematic error, naturally the technical aspect is more important, since a small systematic error is only obtained within sophisticated approximations. 
Put differently, the best formulation does not help if it cannot be implemented beyond a basic truncation. 
Moreover, as eluded to in \Cref{sec:GaugeInvariant+Composites}, gauge invariance in a flow may come with the price of an inherent non-locality. 
The latter is far more cumbersome in terms of a systematic error control than using a gauge-fixed formulation: First of all, non-local flows show generically bad convergence properties for vertex and (covariant) derivative expansions, see \cite{Ihssen:2024miv}. 
Second of all, one may absorb relevant physics into field transformations. 
Here, a relevant example is the confinement mechanism in covariant gauges. In a gauge-fixed \textit{local}  approach, confinement emerges from the Schwinger mechanism that exhibits a massless excitation in the ghost-gluon sector of QCD, for a recent review see \cite{Ferreira:2025anh}. 
Absorbing this part of the dynamics in a non-local redefinition of the gauge field in a quest for gauge invariance is potentially dangerous. However, it might also be a bliss and simplify the non-trivial coupled dynamics. Another example is given by quantum gravity in a gauge invariant single-metric formulation and its intricacies discussed in \Cref{sec:OneLoopGravity-PIG}. 

In conclusion, the quest for best-suited flows in gauge theories has only been started yet and we consider the recent developments as very promising. The PIRG approach to gauge theories connects all different proposals and its flexibility certainly helps in this quest.

%%%%%%%%%%%%%%%%%%%%%%%%%%%%%%%%%%%%%%%%
\section{Conclusion} 
\label{sec:Conclusion}

We have derived general gauge and BRST-invariant flows in the physics-informed functional renormalisation group (PIRG). At the root of this approach is the novel perspective on fRG flows by 
viewing them as flows for the pair $(\Gamma_\phi,\dot\phi)$ with the target action $\Gamma_\phi$ and the the flowing field $\phi$, defined by the respective general field transformation $\dot\phi$. In gauge theories we use the flowing field $\dot \phi$ to absorb all modifications of BRST and gauge invariance, as well as to guarantee a trivial Nielsen identity, that is a single-field formulation, in each RG step. 
The final setup is provided by \labelcref{eq:PIG} in \Cref{sec:GaugeINvariantPIRGS} with the gauge and BRST invariant flow \labelcref{eq:PIGflow} and the accompanying flowing field \labelcref{eq:FlowNielsenSolutionRecall}. 

One of the immediate applications is the construction of flowing fields that elevate the background field approximation of flows in gauge theories to an exact flow. This approximation is commonly used in gravity applications and also in many applications to non-Abelian gauge theories. 
The application of the gauge invariant PIRG at one-loop and beyond has been addressed in \Cref{sec:OneLoopPIGs}. We have used the gauge invariant PIRG in \Cref{sec:OneLoopYM-PIG} to resolve a well-known deficiency of the background field approximation, namely the failure to capture the one-loop $\beta$-function with singular regulators. 
Moreover, we have discussed minimal formulations of gauge invariant flows in which the flowing field carries no part of the relevant operators of the theory. 

In \Cref{sec:OneLoopGravity-PIG} we have discussed the application of gauge invariant PIRGs to gravity. The current approach can be used to resolve two known deficiencies of the background field approach to gravity: the qualitatively different flow and fixed point equation of the graviton mass parameter and the cosmological constant, and the singularity of the background field approximation for $\lambda=1/2$, where $\lambda=\Lambda/k^2$ is the cosmological constant $\Lambda$ measured in units of the cutoff $k$. 

Moreover, we have discussed the landscape of flows in gauge theories through the lens of the novel PIRG approach to gauge theories in \Cref{sec:GaugeInvariantLandscape}, and we refer the reader to the discussion there. 

The applications of the gauge invariant PIRGs go far beyond these examples and we hope to report on some of them in the near future.

%%%%%%%%%%%%%%%%%%%%%%%%%%
\begin{acknowledgements}
We thank Holger Gies, Benjamin Knorr, Silas Mezger, Manuel Reichert, Paul Sprenger and Christof Wetterich for discussions and work on related subjects. 
This work is funded by the Deutsche Forschungsgemeinschaft (DFG, German Research Foundation) under Germany’s Excellence Strategy EXC 2181/1 - 390900948 (the Heidelberg STRUCTURES Excellence Cluster) and the Collaborative Research Centre SFB 1225 (ISOQUANT). It is also supported by EMMI. 
\end{acknowledgements}
%%%%%%%%%%%%%%%%%%%%%%%%	

%%%%%%%%%%%%%%%%%%%%
\appendix

%%%%%%%%%%%%%%%%%%%%%%%%%%%
\section{Background field approach to Yang-Mills theory}
\label{app:BackgroundYM}

In this Appendix we summarise the background field approach to Yang-Mills theory and the standard fRG setup within the background field approach. While we use both Yang-Mills theory and gravity as examples, we perform the explicit illustrative computations only in Yang-Mills theory. For more details and a survey of the results in non-Abelian gauge theories see \cite{Dupuis:2020fhh}.

%%%%%%%%%%%%%%%%%%%%%%%%%%%%%%%%%%%%%
\subsection{Gauge-fixed Yang-Mills theory}

The classical action of Yang-Mills theory is given by
\begin{subequations}\label{eq:ClassAct}
\begin{align}
		S_A[A] = \frac14 \int F^a_{\mu\nu} F^a_{\mu\nu} \,,
\label{eq:Sclassical}
\end{align}
with the fieldstrength 
\begin{align}
		F^a_{\mu\nu} = \partial_\mu A^a_\nu - \partial_\nu A^a_\mu + g f^{abc} A^b_\mu A^c_\nu \,.
\end{align}
\end{subequations}
The trace is taken over the fundamental representation of the gauge group SU($N_c$), with $\tr (t^a t^b) = \frac12 \delta^{ab}$, $a,b = 1, \dots , N_c^2 -1$ and the covariant derivative is given by 
\begin{align}
	D_\mu = \partial_\mu - \imag g_s\, A_\mu\,, 
	\label{eq:CovariantDerivative}
\end{align}
and hence that in the adjoint representation has the form 
\begin{align}
	D_\mu^{bc} = \delta^{bc}\partial_\mu - g_s\, A^a_\mu f^{abc}\,, 
	\label{eq:CovDerAd}
\end{align}
with $(t^a)^{bc} = -i \, f^{abc}$. The action is invariant under infinitesimal gauge transformations $\delta_\omega$, derived from the gauge transformation $U = \exp( i \omega)$, i.e.
\begin{align}
	\delta_\omega S_A[A] = 0 \,, 
\end{align}
with 
\begin{align}
	\delta_\omega A_\mu = \frac{1}{g_s} D_\mu \omega \,.
	\label{eq:GaugeTrafos}
\end{align}
We recall the standard quantised and gauge-fixed action of Yang-Mills theory in the general covariant gauge,
\begin{subequations}
\label{eq:Sgaugefixed}
\begin{align}
		S_\textrm{YM} [A,c,\bar c] = S_A[A] + S_\textrm{gf}[A]+S_\textrm{gh}[A,c,\bar c] \,,
		\label{eq:Sgaugefixedsum} 
\end{align}
with the gauge fixing and ghost actions 
\begin{align}\nonumber 
		S_\textrm{gf}[A] =&\, \frac{1}{2 \xi} \int \left(\partial_\mu A_\mu^a\right)^2 \,,\\[1ex]
		S_\textrm{gh}[A,c,\bar c] =&\, - \int \bar c^a \partial_\mu D^{ab}_\mu c^b \,. 
		\label{eq:Sgf+Sgh}
\end{align}
The covariant gauge fixing condition reads  
\begin{align}
		\partial_\mu A^a_\mu = 0 \,,
\end{align}
\end{subequations}
where $\xi \to 0$ corresponds the Landau gauge. The ghost and anti-ghost $\bar c , c$ are Grassmann valued fields. 
Note that we have chosen a positive sign for the ghost action \labelcref{eq:Sgf+Sgh}, the common choice comes with a relative minus sign. 

We emphasise that the covariant gauge was chosen for the sake of convenience and for working within a specific important example. The gauge invariant PIRG approach, however, is general and applies to all gauges and gauge theories.

%%%%%%%%%%%%%%%%%%%%%%%%%
\subsection{Background field approach}
\label{app:BGgauge}

The background field method admits a gauge invariant effective action, see e.g.~\cite{Abbott:1980hw}. This is achieved by splitting the full gauge field $A_\mu$ into a background field $\bar A_\mu$ and a fluctuation field $a_\mu$,
\begin{align}
	A_\mu = \bar A_\mu +a_\mu \,. 
	\label{eq:split}
\end{align}
Accordingly, the respective effective action $\Gamma$ now depends on both fields, 
\begin{align}
	\Gamma=\Gamma[\bar A,a,c,\bar c]\,.
	\label{eq:GammaBack}
\end{align}
The Yang-Mills action \labelcref{eq:Sgaugefixed} only depends on the sum of the fields \labelcref{eq:split}, but would not be gauge invariant. A gauge invariant effective action can be obtained with the background field gauge. The price to pay is a genuine dependence on the background field that originates from this gauge fixing condition,  
\begin{align}
	\bar D_\mu a_\mu = 0 \,,
	\label{eq:BackgroundGaugeFixing}
\end{align}
where $\bar D_\mu = D_\mu(\bar A)$ is the background covariant derivative \labelcref{eq:CovariantDerivative}. This leads to the classical gauge-fixed action of the background field approach, 
\begin{subequations} 
\label{sec:ActionYMBack} 
\begin{align}
	S_\textrm{YM}[\bar A,\Phi_f] =S_A[\bar A+a]+  S_\textrm{gf}[\bar A, a, b]+ S_\textrm{gh}[\bar A,c,\bar c]\,, 
	\label{sec:SYMBack}
\end{align}
with the gauge fixing and ghost terms, 
\begin{align}\nonumber 
	S_\textrm{gf}[\bar A, a, b]=& \int \textrm{d}^{d} x\left[-\frac{\xi}{2} b^{a} b^{a}+b^{a} \left( \bar D_\mu a_\mu\right)^a\right]\,, \\[1ex]
	S_\textrm{gh}[\bar A,c,\bar c]=&\, - \int \bar c^a \bar D_\mu D^{ab}_\mu c^b \,, 
	\label{eq:BackgroundSgf+SghYM}
\end{align}
\end{subequations} 
where we have also introduced the Nakanishi-Lautrup field $b$. This is simply \labelcref{eq:Sgf+Sgh} with $\partial\to \bar D$. In \labelcref{sec:SYMBack} we have also introduced the fluctuation superfield of the fundamental fields, 
\begin{align}
	\Phi_f=(a, c,\bar c, b)\,. 
	\label{eq:Superfieldf}
\end{align}
%

%%%%%%%%%%%%%%%%%%%%%%
\subsection{Gauge and BRST-invariance}
\label{eq:BRST+Gauge}

\Cref{sec:SYMBack} is invariant under the BRST transformations, $\delta_\varepsilon\, S_\textrm{YM}[\bar A,\Phi] =0$ with   
\begin{align}
 s_\varepsilon A=\varepsilon D c\,, \ \ 
  s_\varepsilon c=\varepsilon i g c^2\,, \ \ 
s_\varepsilon \bar{c}=\varepsilon b\,,  \ \ 
s_\varepsilon b =0\,.
	\label{eq:BRST_Transformations}
\end{align}
Classical BRST-invariance translates into the quantum BRST-invariance of the full effective action, conveniently written in the quantum master equation, 
\begin{align}
	 \frac{\delta \Gamma[\bar A,\Phi_f]}{\delta Q_{i}}\frac{\delta \Gamma[\bar A,\Phi_f]}{\delta \Phi_{f,i}} =0 \,,
	\label{eq:MasterBRST}
\end{align}
where the $Q_i$ are the BRST currents of the fields $\Phi_{f,i}$. 
In \labelcref{eq:MasterBRST} we have used DeWitt's condensed notation and the index $a$ now also includes an integral over space-time. In the presence of the regulator, 	\labelcref{eq:MasterBRST} turns into the modified master equation, 
\begin{align}
	\frac{\delta \Gamma_k}{\delta Q_{f,i}}  \frac{\delta \Gamma_k}{\delta \Phi_{f,i}}
	=
	R^{ij} G_{jl}  \frac{\delta^2 \Gamma_k[A,\Phi_f]}{\delta\Phi_{f,l} \delta Q^i}
	\, ,
	\label{eq:mMaster}
\end{align}
and the right-hand side comprises the breaking of BRST-invariance of the cutoff term. At vanishing cutoff it vanishes and \labelcref{eq:mMaster} reduces to \labelcref{eq:MasterBRST}. 

The gauge-fixed action 	\labelcref{sec:ActionYMBack}  is also invariant under the background gauge transformations $\bar \delta_\omega$ with  
\begin{subequations} 
\label{eq:BackgroundGaugeTrafos}
\begin{align}
		\bar \delta_\omega a_\mu = i [\omega, a_\mu], \qquad 	\bar \delta_\omega \bar{A}_\mu = \frac{1}{g_s} \bar D_\mu \omega\,, 
		\label{eq:BackgroundGaugeTrafos-abarA}
\end{align}
for the background field $\bar A$ and the fluctuation field $a$. The ghost and anti-ghost transform as 
\begin{align}
		\bar \delta_\omega c = i [\omega, c]\,,\qquad  \bar \delta_\omega \bar c = i [\omega, \bar c]\,.
\label{eq:BackgroundGaugeTrafos-cbarc}
\end{align}
\end{subequations} 
Evidently, $S_\textrm{gf}[\bar A,a]$ and $S_\textrm{gh}[\bar A,\Phi]$ are invariant under the background gauge transformations \labelcref{eq:BackgroundGaugeTrafos}, as is the classical Yang-Mills action \labelcref{eq:Sclassical} with 
\begin{align} 
	\bar \delta _\omega A_\mu = \frac{1}{g_s} D_\mu \omega =\delta_\omega A_\mu\,. 
\end{align}
This leads to the background gauge invariance of the full effective action $\Gamma[\bar A, a]$ even in the presence of the regulator, if the latter transforms with 
\begin{align} 
	\bar \delta_\omega R_k = i [\omega, R_k]\,. 
\end{align} 
We conclude that 
\begin{align}
	\bar\delta_\omega \Gamma_k[\bar A, \Phi_f]=0\,, 
	\label{eq:GaugeINvEffActionabarA}
\end{align}
where the effective action is defined as the Legendre transform with respect to the fluctuation fields \labelcref{eq:Superfieldf}. 
\Cref{eq:GaugeINvEffActionabarA} entails that we can readily define a gauge invariant effective action $\Gamma_k[A]$ with 
\begin{align} 
	\Gamma_k[A]= \Gamma_k[A,a=0]\,, 
\end{align}
where we dropped the dependences of the ghosts and the Nakanishi-Lautrup field. 
Note that background gauge invariance has been introduced in \labelcref{eq:BackgroundGaugeTrafos} as an \textit{auxiliary} gauge symmetry and 
such a construction even works for a theory which is not gauge invariant to begin with. \textit{Physical} or quantum gauge symmetry is still carried by the gauge transformations $\delta_\omega$ defined in \labelcref{eq:GaugeTrafos} for the full dynamical gauge field $A$. 
For the full quantum theory it is a symmetry under transformations of the path integral measure, which is that of the fluctuation field: $d a$. 
Hence, \labelcref{eq:GaugeTrafos} is provided by a full gauge transformation of the fluctuation field and the trivial transformation of the background field, 
\begin{align}
	\delta_\omega a_\mu =  \frac{1}{g_s} D_\mu \omega\,,\qquad  \delta_\omega \bar A_\mu=0\,.
	\label{eq:QuantumGaugeTrafo}
\end{align}
\Cref{eq:QuantumGaugeTrafo} is not an invariance of the gauge-fixed action $S_A[\bar A,\Phi_f]$, which is invariant under the respective BRST transformations \labelcref{eq:BRST_Transformations}, with $\omega\to c$ in \labelcref{eq:QuantumGaugeTrafo}.

%%%%%%%%%%%%%%%%%%%%%
\subsection{Nielsen identity}
\label{app:Nielsen} 
	
The Nielsen identity ensures the background independence of physical observables. It is derived by considering the Dyson-Schwinger equation for the difference between background derivatives and fluctuation derivatives. 
In this Appendix we consider Yang-Mills theory for the explicit derivations, the respective one in gravity is structurally identical.  
In the standard background field approach the Nielsen identity takes the form  
\begin{align} 
	\left[	\frac{\delta}{\delta \bar A} - \frac{\delta}{\delta a}\right] \Gamma[\bar A, \Phi_f] 
 =\left \langle  \left[	\frac{\delta}{\delta \bar A} - \frac{\delta}{\delta \hat a}\right] S_\textrm{gauge}[\bar A, \hat \Phi_f]\right\rangle \,, 
	\label{eq:NIy} 
\end{align}
where $\hat \Phi_f$ is the fluctuation field operator and $\Phi_f=\langle \hat \Phi_f\rangle$. \Cref{eq:NIy} has the form of a Master equation, as its right-hand side is related to the BRST variations. 
In the present work we are only interested in the structure of \labelcref{eq:NIy}: if the right-hand side is vanishing, derivatives of the fluctuation field are the same as those of the background field. 
This would entail $\Gamma[\bar A,a] = \Gamma[\bar A+a] $ and the effective action would be gauge invariant due to \labelcref{eq:GaugeINvEffActionabarA}. 
However, the gauge fixing was introduced in the first place to remove the redundancy of the gauge field and necessarily we have $\Gamma[\bar A,a] \neq \Gamma[\bar A+a] $. 
Still, we may hope for a reformulation, where this necessary condition is 'minimal' and is restricted to the classical term. 
For the following discussion we rewrite the effective action in its classical gauge fixing part and the rest. 
Since the rest accommodates all quantum fluctuations, we call it $\Gamma_\textrm{qu}$, even though it also includes the classical Yang-Mills action $S_A[\bar A+a]$, 
\begin{align}
	\Gamma_\textrm{qu}[\bar A,\Phi_f]= \Gamma[\bar A,\Phi]- S_\textrm{gf}[\bar A, \hat a, \hat b]- S_\textrm{gh}[\bar A,\hat c,\hat{\bar c}]\,. 
\end{align}
Then, \labelcref{eq:NIy} takes the form 
\begin{align}
\textrm{NI}[\bar\varphi,\Phi_f]= 	\left[	\frac{\delta}{\delta \bar A} - \frac{\delta}{\delta a}\right] \Gamma_\textrm{qu}[\bar A, \Phi_f]  -\textrm{NI}_\textrm{diag}[\bar A,\Phi_f]=0 \,, 
	\label{eq:NIloop} 
\end{align}
where $\textrm{NI}_\textrm{diag}$ comprises the diagrammatic contributions to the right-hand side of \labelcref{eq:NIy}, 
\begin{align} 
\textrm{NI}_\textrm{diag} = \left \langle  \left[	\frac{\delta}{\delta \bar A} - \frac{\delta}{\delta \hat a}\right] S_\textrm{gauge}\right\rangle - S_\textrm{gauge}[\bar A,\Phi_f]\,.
\end{align}
In the presence of the regulator, \labelcref{eq:NIy} turns into the modified Nielsen identity with a structure that is similar to the modified master equation \labelcref{eq:mMaster}. It reads 
\begin{align}
\mathrm{NI}[\bar A,\Phi_f] = \frac12  \Tr \frac{\delta R_a}{\delta \bar A} G_{aa} + \Tr \frac{\delta R_c}{\delta \bar A} G_{c \bar c} \,, 
\label{eq:modNielsenID}
\end{align}
where the left hand side is the standard Nielsen identity \labelcref{eq:NIloop}. Hence, for vanishing regulator, \labelcref{eq:modNielsenID} turns into the standard Nielsen identity. 
For more details see \cite{Dupuis:2020fhh, Pawlowski:2020qer} and references therein.

%%%%%%%%%%%%%%%%%%%%%%%%%%%%%%%%%%%%%%
\subsection{Standard background field fRG at one-loop} 
\label{app:StandardBack1loop}

The one-loop effective action $\Gamma^{\textrm{1-loop}}[\bar A, a]$ can be written concisely as 
\begin{align}
	\Gamma^{\textrm{1-loop}}[\bar A, a] = 	\frac{1}{2}\int_x \tr F^2\left( Z^{1/2}_F\bar A+ Z_a^{1/2} a \right) + \cdots \,. 
	\label{eq:Gammaa1loop} 
\end{align}
\Cref{eq:Gammaa1loop} reduces to the background field effective action $ \Gamma[\bar A]$, \labelcref{eq:GammaBack} at $a=0$ with 
\begin{align} 
	\Gamma^{\textrm{1-loop}}[A] = 	\frac{Z_F}{2}\int_x \tr F^2\left(A\right) + \cdots \,, 
	\label{eq:GBack1loop}
\end{align} 
and the coupling $Z_F^{1/2} g_s$. The effective action in \labelcref{eq:GBack1loop} is background gauge invariant. Moreover, we have the fluctuation field vertices 
\begin{align} 
	\Gamma^{(3)}_{a^n} = Z_a^{n/2} S_A^{(n)} +\cdots \,.
	\label{eq:Gan1loop}
\end{align}
The vertices \labelcref{eq:Gan1loop} satisfy the Slavnov-Taylor identities, that is the master equation \labelcref{eq:MasterBRST}. 
The difference between the background vertices derived from \labelcref{eq:GBack1loop} and the fluctuations ones in \labelcref{eq:Gan1loop} originate in $S_\textrm{gauge}$ which has a different dependence on $\bar A$ and on $a$. 
Moreover, the combination $(\bar A + Z_a^{1/2}/Z_F^{1/2} a)$ serves as the field $\bar\varphi + \phi$ with a trivial flowing field $\dot\phi= - \frac{\eta_a}{2} \phi$. This entails $\phi= Z_a^{1/2} a$ and $\eta_a =\partial_t \log Z_a $. 

In the presence of the regulator $R_k(\bar A)$ in the background field approach, the one-loop effective action also receives contributions that originate in the background field dependence of the regulator. In particular, a term proportional to $\int_x \tr F^2(\bar A)$ is generated. We have 
\begin{align}
	\Gamma^{\textrm{1-loop}}[\bar A, a] = 	\frac{1}{2}\int_x \tr F^2+ \frac{1}{2}\int_x \tr F^2(\bar A)+ \cdots \,, 
	\label{eq:Gammaa1loopReg} 
\end{align}
where the first term stands for \labelcref{eq:Gammaa1loop}. For the sake of simplicity we have dropped mixed terms with $a$ and the background field $\bar A$ from the regulator.

%%%%%%%%%%%%%%%%%%%%%%%%%%%%%%%%%%%%%%%%
\section{Background field approach to gravity}
\label{app:BackgroundGravity}

In this Appendix we discuss the classical gauge-fixed action of Einstein-Hilbert gravity with a background gauge fixing, for a detailed introduction adjusted to the present applications see \cite{Pawlowski:2020qer}. The full gauge-fixed gravity action is given by 
\begin{align}
	S_\textrm{grav}[\bar g, h] = S^{\ }_\text{EH}[\bar g+ h ] +S_{\textrm{gauge}}[\bar g, \Phi_f]\,, 
\end{align}
with the fundamental gravity superfield 
\begin{align}
	\Phi_f=( h_{\mu\nu}, c_\mu, \bar c_\mu) \,.
\label{eq:PhifGravity} 
\end{align}
The classical action is given by 
\begin{align} 
	S^{\ }_\text{EH}[g_{\mu\nu}]= \frac{1}{16 \pi G_N} \int\! \mathrm d^4 x \sqrt{g} \left(2 \Lambda - R\right)\,,
		\label{eq:EH-Action} 
\end{align}
with the abbreviation $g= \text{det}\,g_{\mu\nu}$. \Cref{eq:EH-Action} depends on the couplings $G_N$ (Newton constant) and $\Lambda$ (cosmological constant). The curvature scalar is given by $R= R^{\mu\nu}{}_{\mu\nu} $ with the curvature tensor $R^\rho{}_{\sigma\mu\nu}$ and the Levi-Civita connection $\Gamma$ with 
\begin{align}\nonumber 
		R^\rho{}_{\sigma\mu\nu} =&\,  \partial_\mu \Gamma^\rho{}_{\nu\sigma} -
		\partial_\nu \Gamma^\rho{}_{\mu\sigma} +
		\Gamma^\rho{}_{\mu\lambda}\Gamma^\lambda{}_{\nu\sigma} -
		\Gamma^\rho{}_{\nu\lambda}\Gamma^\lambda{}_{\mu\sigma} \, , \\[1ex]
		\Gamma^\sigma{}_{\mu\nu}=&\,  \tfrac{1}{2}g^{\sigma\rho}\Bigl( \partial_\mu g_{\nu\rho} +
		\partial_\nu g_{\rho\mu} - \partial_\rho g_{\mu\nu} \Bigr) \,.  
\label{eq:Curvature+LeviCivita}
\end{align}
The classical gauge fixing sector is given by  
\begin{align} 
	S_\textrm{\textrm{gauge}}[\bar g, 	\Phi_f] = S_\textrm{gf}[\bar g, h]+ S_\textrm{gh}[\bar g,\Phi_f]\,.  
\end{align} 
Analogously to Yang-Mills theory we introduce a background gauge fixing sector. In contradistinction to the latter, a linear gauge fixing (quadratic in the fluctuation field $h_{\mu\nu}$) requires already the introduction of a background in the space-time volume $d^4 x \,\sqrt{\bar g}$. A general gauge fixing  reads 
\begin{subequations} 
	\label{eq:SgaugeGravity} 
\begin{align}
	\label{eq:gf} 
	S_{\text{gf}}[\bar g, h]=\frac{1}{2 \alpha} \int \!\mathrm{d}^4x
	\sqrt{\bar{g}}\; \bar{g}^{\mu \nu} F_\mu F_\nu \,, 
\end{align}
and linear gauge fixings are linear in the fluctuation field $h_{\mu\nu}$. The linear background gauge fixing condition $F_\mu$ is given by
\begin{align}
	\label{eq:gf-condition}
	F_\mu[\bar g, h] =
	\bar{\nabla}^\nu h_{\mu \nu} -\frac{1+ \beta}{4} \bar{\nabla}_\mu h^{\nu}_{~\nu} \,, 
\end{align}
where $\bar\nabla$ is the covariant derivative with the background metric $\bar g_{\mu\nu}$. 

The ghost action related to \labelcref{eq:gf-condition} reads 
\begin{align}
	\label{eq:Sghost}
	S_{\text{gh}}[\bar g, \phi]=\int \!\mathrm{d}^4x
	\sqrt{\bar{g}}\; \bar c^\mu M_{\mu\nu} c^\nu\,, 
\end{align}
with the Faddeev-Popov operator 
\begin{align}
	\label{eq:OpFP}
	M_{\mu\nu}= \bar\nabla^\rho\! \left(g_{\mu\nu} \nabla_\rho +g_{\rho\nu} \nabla_\mu\right) -\frac{1+\beta}{2} \bar g^{\sigma\rho} \bar\nabla_\mu g_{\nu\sigma} \nabla_\rho\,.
\end{align}
\end{subequations} 
Again, $\bar\nabla$ is the covariant derivative with the background metric $\bar g_{\mu\nu}$ while $\nabla$ is that with the full metric $g_{\mu\nu}$. 
Note that $M_{\mu\nu}$ is linear in the fluctuation field $h$. 
For a detailed discussion of the (modified) Nielsen identity, the (modified) master equations and their relation to gauge invariance and background gauge invariance we refer to \cite{Pawlowski:2020qer}, Chapter VI.

\begingroup
\allowdisplaybreaks
%%%%%%%%%%%%%%%%%%%%%%%%%%%%%%%%%%
	
	\bibliographystyle{apsrev4-2}
	\bibliography{references}

%apsrev4-2.bst 2019-01-14 (MD) hand-edited version of apsrev4-1.bst
%Control: key (0)
%Control: author (72) initials jnrlst
%Control: editor formatted (1) identically to author
%Control: production of article title (-1) disabled
%Control: page (0) single
%Control: year (1) truncated
%Control: production of eprint (0) enabled
\begin{thebibliography}{73}%
\makeatletter
\providecommand \@ifxundefined [1]{%
 \@ifx{#1\undefined}
}%
\providecommand \@ifnum [1]{%
 \ifnum #1\expandafter \@firstoftwo
 \else \expandafter \@secondoftwo
 \fi
}%
\providecommand \@ifx [1]{%
 \ifx #1\expandafter \@firstoftwo
 \else \expandafter \@secondoftwo
 \fi
}%
\providecommand \natexlab [1]{#1}%
\providecommand \enquote  [1]{``#1''}%
\providecommand \bibnamefont  [1]{#1}%
\providecommand \bibfnamefont [1]{#1}%
\providecommand \citenamefont [1]{#1}%
\providecommand \href@noop [0]{\@secondoftwo}%
\providecommand \href [0]{\begingroup \@sanitize@url \@href}%
\providecommand \@href[1]{\@@startlink{#1}\@@href}%
\providecommand \@@href[1]{\endgroup#1\@@endlink}%
\providecommand \@sanitize@url [0]{\catcode `\\12\catcode `\$12\catcode
  `\&12\catcode `\#12\catcode `\^12\catcode `\_12\catcode `\%12\relax}%
\providecommand \@@startlink[1]{}%
\providecommand \@@endlink[0]{}%
\providecommand \url  [0]{\begingroup\@sanitize@url \@url }%
\providecommand \@url [1]{\endgroup\@href {#1}{\urlprefix }}%
\providecommand \urlprefix  [0]{URL }%
\providecommand \Eprint [0]{\href }%
\providecommand \doibase [0]{https://doi.org/}%
\providecommand \selectlanguage [0]{\@gobble}%
\providecommand \bibinfo  [0]{\@secondoftwo}%
\providecommand \bibfield  [0]{\@secondoftwo}%
\providecommand \translation [1]{[#1]}%
\providecommand \BibitemOpen [0]{}%
\providecommand \bibitemStop [0]{}%
\providecommand \bibitemNoStop [0]{.\EOS\space}%
\providecommand \EOS [0]{\spacefactor3000\relax}%
\providecommand \BibitemShut  [1]{\csname bibitem#1\endcsname}%
\let\auto@bib@innerbib\@empty
%</preamble>
\bibitem [{\citenamefont {Ihssen}\ and\ \citenamefont
  {Pawlowski}(2024)}]{Ihssen:2024ihp}%
  \BibitemOpen
  \bibfield  {author} {\bibinfo {author} {\bibfnamefont {F.}~\bibnamefont
  {Ihssen}}\ and\ \bibinfo {author} {\bibfnamefont {J.~M.}\ \bibnamefont
  {Pawlowski}},\ }\href@noop {} {\  (\bibinfo {year} {2024})},\ \Eprint
  {https://arxiv.org/abs/2409.13679} {arXiv:2409.13679 [hep-th]} \BibitemShut
  {NoStop}%
\bibitem [{\citenamefont {Ihssen}\ and\ \citenamefont
  {Pawlowski}(2023{\natexlab{a}})}]{Ihssen:2023nqd}%
  \BibitemOpen
  \bibfield  {author} {\bibinfo {author} {\bibfnamefont {F.}~\bibnamefont
  {Ihssen}}\ and\ \bibinfo {author} {\bibfnamefont {J.~M.}\ \bibnamefont
  {Pawlowski}},\ }\href@noop {} {\  (\bibinfo {year} {2023}{\natexlab{a}})},\
  \Eprint {https://arxiv.org/abs/2305.00816} {arXiv:2305.00816 [hep-th]}
  \BibitemShut {NoStop}%
\bibitem [{\citenamefont {Pawlowski}(2007)}]{Pawlowski:2005xe}%
  \BibitemOpen
  \bibfield  {author} {\bibinfo {author} {\bibfnamefont {J.~M.}\ \bibnamefont
  {Pawlowski}},\ }\href {https://doi.org/10.1016/j.aop.2007.01.007} {\bibfield
  {journal} {\bibinfo  {journal} {Annals Phys.}\ }\textbf {\bibinfo {volume}
  {322}},\ \bibinfo {pages} {2831} (\bibinfo {year} {2007})},\ \Eprint
  {https://arxiv.org/abs/hep-th/0512261} {arXiv:hep-th/0512261} \BibitemShut
  {NoStop}%
\bibitem [{\citenamefont {Litim}\ and\ \citenamefont
  {Pawlowski}(2002{\natexlab{a}})}]{Litim:2002xm}%
  \BibitemOpen
  \bibfield  {author} {\bibinfo {author} {\bibfnamefont {D.~F.}\ \bibnamefont
  {Litim}}\ and\ \bibinfo {author} {\bibfnamefont {J.~M.}\ \bibnamefont
  {Pawlowski}},\ }\href {https://doi.org/10.1103/PhysRevD.66.025030} {\bibfield
   {journal} {\bibinfo  {journal} {Phys. Rev. D}\ }\textbf {\bibinfo {volume}
  {66}},\ \bibinfo {pages} {025030} (\bibinfo {year} {2002}{\natexlab{a}})},\
  \Eprint {https://arxiv.org/abs/hep-th/0202188} {arXiv:hep-th/0202188}
  \BibitemShut {NoStop}%
\bibitem [{\citenamefont {Bergner}\ \emph {et~al.}(2013)\citenamefont
  {Bergner}, \citenamefont {Bruckmann}, \citenamefont {Echigo}, \citenamefont
  {Igarashi}, \citenamefont {Pawlowski},\ and\ \citenamefont
  {Schierenberg}}]{Bergner:2012nu}%
  \BibitemOpen
  \bibfield  {author} {\bibinfo {author} {\bibfnamefont {G.}~\bibnamefont
  {Bergner}}, \bibinfo {author} {\bibfnamefont {F.}~\bibnamefont {Bruckmann}},
  \bibinfo {author} {\bibfnamefont {Y.}~\bibnamefont {Echigo}}, \bibinfo
  {author} {\bibfnamefont {Y.}~\bibnamefont {Igarashi}}, \bibinfo {author}
  {\bibfnamefont {J.~M.}\ \bibnamefont {Pawlowski}},\ and\ \bibinfo {author}
  {\bibfnamefont {S.}~\bibnamefont {Schierenberg}},\ }\href
  {https://doi.org/10.1103/PhysRevD.87.094516} {\bibfield  {journal} {\bibinfo
  {journal} {Phys. Rev. D}\ }\textbf {\bibinfo {volume} {87}},\ \bibinfo
  {pages} {094516} (\bibinfo {year} {2013})},\ \Eprint
  {https://arxiv.org/abs/1212.0219} {arXiv:1212.0219 [hep-lat]} \BibitemShut
  {NoStop}%
\bibitem [{\citenamefont {Gies}(2012)}]{Gies:2006wv}%
  \BibitemOpen
  \bibfield  {author} {\bibinfo {author} {\bibfnamefont {H.}~\bibnamefont
  {Gies}},\ }\href {https://doi.org/10.1007/978-3-642-27320-9_6} {\bibfield
  {journal} {\bibinfo  {journal} {Lect. Notes Phys.}\ }\textbf {\bibinfo
  {volume} {852}},\ \bibinfo {pages} {287} (\bibinfo {year} {2012})},\ \Eprint
  {https://arxiv.org/abs/hep-ph/0611146} {arXiv:hep-ph/0611146} \BibitemShut
  {NoStop}%
\bibitem [{\citenamefont {Braun}(2012)}]{Braun:2011pp}%
  \BibitemOpen
  \bibfield  {author} {\bibinfo {author} {\bibfnamefont {J.}~\bibnamefont
  {Braun}},\ }\href {https://doi.org/10.1088/0954-3899/39/3/033001} {\bibfield
  {journal} {\bibinfo  {journal} {J. Phys. G}\ }\textbf {\bibinfo {volume}
  {39}},\ \bibinfo {pages} {033001} (\bibinfo {year} {2012})},\ \Eprint
  {https://arxiv.org/abs/1108.4449} {arXiv:1108.4449 [hep-ph]} \BibitemShut
  {NoStop}%
\bibitem [{\citenamefont {Dupuis}\ \emph {et~al.}(2021)\citenamefont {Dupuis},
  \citenamefont {Canet}, \citenamefont {Eichhorn}, \citenamefont {Metzner},
  \citenamefont {Pawlowski}, \citenamefont {Tissier},\ and\ \citenamefont
  {Wschebor}}]{Dupuis:2020fhh}%
  \BibitemOpen
  \bibfield  {author} {\bibinfo {author} {\bibfnamefont {N.}~\bibnamefont
  {Dupuis}}, \bibinfo {author} {\bibfnamefont {L.}~\bibnamefont {Canet}},
  \bibinfo {author} {\bibfnamefont {A.}~\bibnamefont {Eichhorn}}, \bibinfo
  {author} {\bibfnamefont {W.}~\bibnamefont {Metzner}}, \bibinfo {author}
  {\bibfnamefont {J.~M.}\ \bibnamefont {Pawlowski}}, \bibinfo {author}
  {\bibfnamefont {M.}~\bibnamefont {Tissier}},\ and\ \bibinfo {author}
  {\bibfnamefont {N.}~\bibnamefont {Wschebor}},\ }\href
  {https://doi.org/10.1016/j.physrep.2021.01.001} {\bibfield  {journal}
  {\bibinfo  {journal} {Phys. Rept.}\ }\textbf {\bibinfo {volume} {910}},\
  \bibinfo {pages} {1} (\bibinfo {year} {2021})},\ \Eprint
  {https://arxiv.org/abs/2006.04853} {arXiv:2006.04853 [cond-mat.stat-mech]}
  \BibitemShut {NoStop}%
\bibitem [{\citenamefont {Fu}(2022)}]{Fu:2022gou}%
  \BibitemOpen
  \bibfield  {author} {\bibinfo {author} {\bibfnamefont {W.-j.}\ \bibnamefont
  {Fu}},\ }\href {https://doi.org/10.1088/1572-9494/ac86be} {\bibfield
  {journal} {\bibinfo  {journal} {Commun. Theor. Phys.}\ }\textbf {\bibinfo
  {volume} {74}},\ \bibinfo {pages} {097304} (\bibinfo {year} {2022})},\
  \Eprint {https://arxiv.org/abs/2205.00468} {arXiv:2205.00468 [hep-ph]}
  \BibitemShut {NoStop}%
\bibitem [{\citenamefont {Pawlowski}\ and\ \citenamefont
  {Reichert}(2021)}]{Pawlowski:2020qer}%
  \BibitemOpen
  \bibfield  {author} {\bibinfo {author} {\bibfnamefont {J.~M.}\ \bibnamefont
  {Pawlowski}}\ and\ \bibinfo {author} {\bibfnamefont {M.}~\bibnamefont
  {Reichert}},\ }\href {https://doi.org/10.3389/fphy.2020.551848} {\bibfield
  {journal} {\bibinfo  {journal} {Front. in Phys.}\ }\textbf {\bibinfo {volume}
  {8}},\ \bibinfo {pages} {551848} (\bibinfo {year} {2021})},\ \Eprint
  {https://arxiv.org/abs/2007.10353} {arXiv:2007.10353 [hep-th]} \BibitemShut
  {NoStop}%
\bibitem [{\citenamefont {Pawlowski}\ and\ \citenamefont
  {Reichert}(2023)}]{Pawlowski:2023gym}%
  \BibitemOpen
  \bibfield  {author} {\bibinfo {author} {\bibfnamefont {J.~M.}\ \bibnamefont
  {Pawlowski}}\ and\ \bibinfo {author} {\bibfnamefont {M.}~\bibnamefont
  {Reichert}},\ }\href@noop {} {\  (\bibinfo {year} {2023})},\ \Eprint
  {https://arxiv.org/abs/2309.10785} {arXiv:2309.10785 [hep-th]} \BibitemShut
  {NoStop}%
\bibitem [{\citenamefont {Litim}\ and\ \citenamefont
  {Pawlowski}(2002{\natexlab{b}})}]{Litim:2002ce}%
  \BibitemOpen
  \bibfield  {author} {\bibinfo {author} {\bibfnamefont {D.~F.}\ \bibnamefont
  {Litim}}\ and\ \bibinfo {author} {\bibfnamefont {J.~M.}\ \bibnamefont
  {Pawlowski}},\ }\href {https://doi.org/10.1088/1126-6708/2002/09/049}
  {\bibfield  {journal} {\bibinfo  {journal} {JHEP}\ }\textbf {\bibinfo
  {volume} {09}},\ \bibinfo {pages} {049}},\ \Eprint
  {https://arxiv.org/abs/hep-th/0203005} {arXiv:hep-th/0203005} \BibitemShut
  {NoStop}%
\bibitem [{\citenamefont {Meibohm}\ \emph {et~al.}(2016)\citenamefont
  {Meibohm}, \citenamefont {Pawlowski},\ and\ \citenamefont
  {Reichert}}]{Meibohm:2015twa}%
  \BibitemOpen
  \bibfield  {author} {\bibinfo {author} {\bibfnamefont {J.}~\bibnamefont
  {Meibohm}}, \bibinfo {author} {\bibfnamefont {J.~M.}\ \bibnamefont
  {Pawlowski}},\ and\ \bibinfo {author} {\bibfnamefont {M.}~\bibnamefont
  {Reichert}},\ }\href {https://doi.org/10.1103/PhysRevD.93.084035} {\bibfield
  {journal} {\bibinfo  {journal} {Phys. Rev. D}\ }\textbf {\bibinfo {volume}
  {93}},\ \bibinfo {pages} {084035} (\bibinfo {year} {2016})},\ \Eprint
  {https://arxiv.org/abs/1510.07018} {arXiv:1510.07018 [hep-th]} \BibitemShut
  {NoStop}%
\bibitem [{\citenamefont {Litim}\ and\ \citenamefont
  {Pawlowski}(2002{\natexlab{c}})}]{Litim:2001ky}%
  \BibitemOpen
  \bibfield  {author} {\bibinfo {author} {\bibfnamefont {D.~F.}\ \bibnamefont
  {Litim}}\ and\ \bibinfo {author} {\bibfnamefont {J.~M.}\ \bibnamefont
  {Pawlowski}},\ }\href {https://doi.org/10.1103/PhysRevD.65.081701} {\bibfield
   {journal} {\bibinfo  {journal} {Phys. Rev. D}\ }\textbf {\bibinfo {volume}
  {65}},\ \bibinfo {pages} {081701} (\bibinfo {year} {2002}{\natexlab{c}})},\
  \Eprint {https://arxiv.org/abs/hep-th/0111191} {arXiv:hep-th/0111191}
  \BibitemShut {NoStop}%
\bibitem [{\citenamefont {Litim}\ and\ \citenamefont
  {Pawlowski}(2002{\natexlab{d}})}]{Litim:2002hj}%
  \BibitemOpen
  \bibfield  {author} {\bibinfo {author} {\bibfnamefont {D.~F.}\ \bibnamefont
  {Litim}}\ and\ \bibinfo {author} {\bibfnamefont {J.~M.}\ \bibnamefont
  {Pawlowski}},\ }\href {https://doi.org/10.1016/S0370-2693(02)02693-X}
  {\bibfield  {journal} {\bibinfo  {journal} {Phys. Lett. B}\ }\textbf
  {\bibinfo {volume} {546}},\ \bibinfo {pages} {279} (\bibinfo {year}
  {2002}{\natexlab{d}})},\ \Eprint {https://arxiv.org/abs/hep-th/0208216}
  {arXiv:hep-th/0208216} \BibitemShut {NoStop}%
\bibitem [{\citenamefont {Pawlowski}(2002)}]{JMP2002}%
  \BibitemOpen
  \bibfield  {author} {\bibinfo {author} {\bibfnamefont {J.~M.}\ \bibnamefont
  {Pawlowski}},\ }\href@noop {} {\bibfield  {journal} {\bibinfo  {journal}
  {Habilitation thesis}\ } (\bibinfo {year} {2002})}\BibitemShut {NoStop}%
\bibitem [{\citenamefont {Folkerts}\ \emph {et~al.}(2012)\citenamefont
  {Folkerts}, \citenamefont {Litim},\ and\ \citenamefont
  {Pawlowski}}]{Folkerts:2011jz}%
  \BibitemOpen
  \bibfield  {author} {\bibinfo {author} {\bibfnamefont {S.}~\bibnamefont
  {Folkerts}}, \bibinfo {author} {\bibfnamefont {D.~F.}\ \bibnamefont
  {Litim}},\ and\ \bibinfo {author} {\bibfnamefont {J.~M.}\ \bibnamefont
  {Pawlowski}},\ }\href {https://doi.org/10.1016/j.physletb.2012.02.002}
  {\bibfield  {journal} {\bibinfo  {journal} {Phys. Lett. B}\ }\textbf
  {\bibinfo {volume} {709}},\ \bibinfo {pages} {234} (\bibinfo {year}
  {2012})},\ \Eprint {https://arxiv.org/abs/1101.5552} {arXiv:1101.5552
  [hep-th]} \BibitemShut {NoStop}%
\bibitem [{\citenamefont {Reuter}\ and\ \citenamefont
  {Wetterich}(1994)}]{Reuter:1993kw}%
  \BibitemOpen
  \bibfield  {author} {\bibinfo {author} {\bibfnamefont {M.}~\bibnamefont
  {Reuter}}\ and\ \bibinfo {author} {\bibfnamefont {C.}~\bibnamefont
  {Wetterich}},\ }\href {https://doi.org/10.1016/0550-3213(94)90543-6}
  {\bibfield  {journal} {\bibinfo  {journal} {Nucl. Phys. B}\ }\textbf
  {\bibinfo {volume} {417}},\ \bibinfo {pages} {181} (\bibinfo {year}
  {1994})}\BibitemShut {NoStop}%
\bibitem [{\citenamefont {Gies}(2002)}]{Gies:2002af}%
  \BibitemOpen
  \bibfield  {author} {\bibinfo {author} {\bibfnamefont {H.}~\bibnamefont
  {Gies}},\ }\href {https://doi.org/10.1103/PhysRevD.66.025006} {\bibfield
  {journal} {\bibinfo  {journal} {Phys. Rev. D}\ }\textbf {\bibinfo {volume}
  {66}},\ \bibinfo {pages} {025006} (\bibinfo {year} {2002})},\ \Eprint
  {https://arxiv.org/abs/hep-th/0202207} {arXiv:hep-th/0202207} \BibitemShut
  {NoStop}%
\bibitem [{\citenamefont {Pernici}\ and\ \citenamefont
  {Raciti}(1998)}]{Pernici:1998tp}%
  \BibitemOpen
  \bibfield  {author} {\bibinfo {author} {\bibfnamefont {M.}~\bibnamefont
  {Pernici}}\ and\ \bibinfo {author} {\bibfnamefont {M.}~\bibnamefont
  {Raciti}},\ }\href {https://doi.org/10.1016/S0550-3213(98)80007-2} {\bibfield
   {journal} {\bibinfo  {journal} {Nucl. Phys. B}\ }\textbf {\bibinfo {volume}
  {531}},\ \bibinfo {pages} {560} (\bibinfo {year} {1998})},\ \Eprint
  {https://arxiv.org/abs/hep-th/9803212} {arXiv:hep-th/9803212} \BibitemShut
  {NoStop}%
\bibitem [{\citenamefont {Rosten}(2012)}]{Rosten:2010vm}%
  \BibitemOpen
  \bibfield  {author} {\bibinfo {author} {\bibfnamefont {O.~J.}\ \bibnamefont
  {Rosten}},\ }\href {https://doi.org/10.1016/j.physrep.2011.12.003} {\bibfield
   {journal} {\bibinfo  {journal} {Phys. Rept.}\ }\textbf {\bibinfo {volume}
  {511}},\ \bibinfo {pages} {177} (\bibinfo {year} {2012})},\ \Eprint
  {https://arxiv.org/abs/1003.1366} {arXiv:1003.1366 [hep-th]} \BibitemShut
  {NoStop}%
\bibitem [{\citenamefont {Ihssen}\ \emph
  {et~al.}(2025{\natexlab{a}})\citenamefont {Ihssen}, \citenamefont {Knorr},
  \citenamefont {Mezger},\ and\ \citenamefont {Pawlowski}}]{IKMP}%
  \BibitemOpen
  \bibfield  {author} {\bibinfo {author} {\bibfnamefont {F.}~\bibnamefont
  {Ihssen}}, \bibinfo {author} {\bibfnamefont {B.}~\bibnamefont {Knorr}},
  \bibinfo {author} {\bibfnamefont {S.}~\bibnamefont {Mezger}},\ and\ \bibinfo
  {author} {\bibfnamefont {J.~M.}\ \bibnamefont {Pawlowski}},\ }\href@noop {}
  {\bibfield  {journal} {\bibinfo  {journal} {in preparation}\ } (\bibinfo
  {year} {2025}{\natexlab{a}})}\BibitemShut {NoStop}%
\bibitem [{\citenamefont {Bambi}\ \emph {et~al.}(2024)\citenamefont {Bambi},
  \citenamefont {Modesto},\ and\ \citenamefont {Shapiro}}]{Bambi2024-nm}%
  \BibitemOpen
  \bibinfo {editor} {\bibfnamefont {C.}~\bibnamefont {Bambi}}, \bibinfo
  {editor} {\bibfnamefont {L.}~\bibnamefont {Modesto}},\ and\ \bibinfo {editor}
  {\bibfnamefont {I.}~\bibnamefont {Shapiro}},\ eds.,\ \href@noop {} {\emph
  {\bibinfo {title} {Handbook of quantum gravity}}}\ (\bibinfo  {publisher}
  {Springer Nature Singapore},\ \bibinfo {address} {Singapore},\ \bibinfo
  {year} {2024})\BibitemShut {NoStop}%
\bibitem [{\citenamefont {Ihssen}\ \emph
  {et~al.}(2025{\natexlab{b}})\citenamefont {Ihssen}, \citenamefont {Knorr},
  \citenamefont {Mezger}, \citenamefont {Pawlowski},\ and\ \citenamefont
  {Sprenger}}]{IKMPS}%
  \BibitemOpen
  \bibfield  {author} {\bibinfo {author} {\bibfnamefont {F.}~\bibnamefont
  {Ihssen}}, \bibinfo {author} {\bibfnamefont {B.}~\bibnamefont {Knorr}},
  \bibinfo {author} {\bibfnamefont {S.}~\bibnamefont {Mezger}}, \bibinfo
  {author} {\bibfnamefont {J.~M.}\ \bibnamefont {Pawlowski}},\ and\ \bibinfo
  {author} {\bibfnamefont {P.}~\bibnamefont {Sprenger}},\ }\href@noop {}
  {\bibfield  {journal} {\bibinfo  {journal} {in preparation}\ } (\bibinfo
  {year} {2025}{\natexlab{b}})}\BibitemShut {NoStop}%
\bibitem [{\citenamefont {Pawlowski}\ \emph {et~al.}(2022)\citenamefont
  {Pawlowski}, \citenamefont {Schneider},\ and\ \citenamefont
  {Wink}}]{Pawlowski:2022oyq}%
  \BibitemOpen
  \bibfield  {author} {\bibinfo {author} {\bibfnamefont {J.~M.}\ \bibnamefont
  {Pawlowski}}, \bibinfo {author} {\bibfnamefont {C.~S.}\ \bibnamefont
  {Schneider}},\ and\ \bibinfo {author} {\bibfnamefont {N.}~\bibnamefont
  {Wink}},\ }\href@noop {} {\  (\bibinfo {year} {2022})},\ \Eprint
  {https://arxiv.org/abs/2202.11123} {arXiv:2202.11123 [hep-th]} \BibitemShut
  {NoStop}%
\bibitem [{\citenamefont {Ihssen}\ \emph {et~al.}(2023)\citenamefont {Ihssen},
  \citenamefont {Sattler},\ and\ \citenamefont {Wink}}]{Ihssen:2023qaq}%
  \BibitemOpen
  \bibfield  {author} {\bibinfo {author} {\bibfnamefont {F.}~\bibnamefont
  {Ihssen}}, \bibinfo {author} {\bibfnamefont {F.~R.}\ \bibnamefont
  {Sattler}},\ and\ \bibinfo {author} {\bibfnamefont {N.}~\bibnamefont
  {Wink}},\ }\href {https://doi.org/10.1103/PhysRevD.107.114009} {\bibfield
  {journal} {\bibinfo  {journal} {Phys. Rev. D}\ }\textbf {\bibinfo {volume}
  {107}},\ \bibinfo {pages} {114009} (\bibinfo {year} {2023})},\ \Eprint
  {https://arxiv.org/abs/2302.04736} {arXiv:2302.04736 [hep-th]} \BibitemShut
  {NoStop}%
\bibitem [{\citenamefont {Sattler}\ and\ \citenamefont
  {Pawlowski}(2024)}]{Sattler:2024ozv}%
  \BibitemOpen
  \bibfield  {author} {\bibinfo {author} {\bibfnamefont {F.~R.}\ \bibnamefont
  {Sattler}}\ and\ \bibinfo {author} {\bibfnamefont {J.~M.}\ \bibnamefont
  {Pawlowski}},\ }\href@noop {} {\  (\bibinfo {year} {2024})},\ \Eprint
  {https://arxiv.org/abs/2412.13043} {arXiv:2412.13043 [hep-ph]} \BibitemShut
  {NoStop}%
\bibitem [{\citenamefont {Fu}\ \emph {et~al.}(2025)\citenamefont {Fu},
  \citenamefont {Huang}, \citenamefont {Pawlowski}, \citenamefont {Tan},\ and\
  \citenamefont {Zhou}}]{Fu:2025hcm}%
  \BibitemOpen
  \bibfield  {author} {\bibinfo {author} {\bibfnamefont {W.-j.}\ \bibnamefont
  {Fu}}, \bibinfo {author} {\bibfnamefont {C.}~\bibnamefont {Huang}}, \bibinfo
  {author} {\bibfnamefont {J.~M.}\ \bibnamefont {Pawlowski}}, \bibinfo {author}
  {\bibfnamefont {Y.-y.}\ \bibnamefont {Tan}},\ and\ \bibinfo {author}
  {\bibfnamefont {L.-j.}\ \bibnamefont {Zhou}},\ }\href@noop {} {\  (\bibinfo
  {year} {2025})},\ \Eprint {https://arxiv.org/abs/2502.14388}
  {arXiv:2502.14388 [hep-ph]} \BibitemShut {NoStop}%
\bibitem [{\citenamefont {Lavrov}\ and\ \citenamefont
  {Shapiro}(2013)}]{Lavrov:2012xz}%
  \BibitemOpen
  \bibfield  {author} {\bibinfo {author} {\bibfnamefont {P.~M.}\ \bibnamefont
  {Lavrov}}\ and\ \bibinfo {author} {\bibfnamefont {I.~L.}\ \bibnamefont
  {Shapiro}},\ }\href {https://doi.org/10.1007/JHEP06(2013)086} {\bibfield
  {journal} {\bibinfo  {journal} {JHEP}\ }\textbf {\bibinfo {volume} {06}},\
  \bibinfo {pages} {086}},\ \Eprint {https://arxiv.org/abs/1212.2577}
  {arXiv:1212.2577 [hep-th]} \BibitemShut {NoStop}%
\bibitem [{\citenamefont {Igarashi}\ \emph {et~al.}(2010)\citenamefont
  {Igarashi}, \citenamefont {Itoh},\ and\ \citenamefont
  {Sonoda}}]{Igarashi:2009tj}%
  \BibitemOpen
  \bibfield  {author} {\bibinfo {author} {\bibfnamefont {Y.}~\bibnamefont
  {Igarashi}}, \bibinfo {author} {\bibfnamefont {K.}~\bibnamefont {Itoh}},\
  and\ \bibinfo {author} {\bibfnamefont {H.}~\bibnamefont {Sonoda}},\ }\href
  {https://doi.org/10.1143/PTPS.181.1} {\bibfield  {journal} {\bibinfo
  {journal} {Prog. Theor. Phys. Suppl.}\ }\textbf {\bibinfo {volume} {181}},\
  \bibinfo {pages} {1} (\bibinfo {year} {2010})},\ \Eprint
  {https://arxiv.org/abs/0909.0327} {arXiv:0909.0327 [hep-th]} \BibitemShut
  {NoStop}%
\bibitem [{\citenamefont {Igarashi}\ \emph {et~al.}(2016)\citenamefont
  {Igarashi}, \citenamefont {Itoh},\ and\ \citenamefont
  {Pawlowski}}]{Igarashi:2016gcf}%
  \BibitemOpen
  \bibfield  {author} {\bibinfo {author} {\bibfnamefont {Y.}~\bibnamefont
  {Igarashi}}, \bibinfo {author} {\bibfnamefont {K.}~\bibnamefont {Itoh}},\
  and\ \bibinfo {author} {\bibfnamefont {J.~M.}\ \bibnamefont {Pawlowski}},\
  }\href {https://doi.org/10.1088/1751-8113/49/40/405401} {\bibfield  {journal}
  {\bibinfo  {journal} {J. Phys. A}\ }\textbf {\bibinfo {volume} {49}},\
  \bibinfo {pages} {405401} (\bibinfo {year} {2016})},\ \Eprint
  {https://arxiv.org/abs/1604.08327} {arXiv:1604.08327 [hep-th]} \BibitemShut
  {NoStop}%
\bibitem [{\citenamefont {Igarashi}\ \emph {et~al.}(2019)\citenamefont
  {Igarashi}, \citenamefont {Itoh},\ and\ \citenamefont
  {Morris}}]{Igarashi:2019gkm}%
  \BibitemOpen
  \bibfield  {author} {\bibinfo {author} {\bibfnamefont {Y.}~\bibnamefont
  {Igarashi}}, \bibinfo {author} {\bibfnamefont {K.}~\bibnamefont {Itoh}},\
  and\ \bibinfo {author} {\bibfnamefont {T.~R.}\ \bibnamefont {Morris}},\
  }\href {https://doi.org/10.1093/ptep/ptz099} {\bibfield  {journal} {\bibinfo
  {journal} {PTEP}\ }\textbf {\bibinfo {volume} {2019}},\ \bibinfo {pages}
  {103B01} (\bibinfo {year} {2019})},\ \Eprint
  {https://arxiv.org/abs/1904.08231} {arXiv:1904.08231 [hep-th]} \BibitemShut
  {NoStop}%
\bibitem [{\citenamefont {Gies}\ \emph {et~al.}(2022)\citenamefont {Gies},
  \citenamefont {Gkiatas},\ and\ \citenamefont {Zambelli}}]{Gies:2022mar}%
  \BibitemOpen
  \bibfield  {author} {\bibinfo {author} {\bibfnamefont {H.}~\bibnamefont
  {Gies}}, \bibinfo {author} {\bibfnamefont {D.}~\bibnamefont {Gkiatas}},\ and\
  \bibinfo {author} {\bibfnamefont {L.}~\bibnamefont {Zambelli}},\ }\href
  {https://doi.org/10.1103/PhysRevD.106.116013} {\bibfield  {journal} {\bibinfo
   {journal} {Phys. Rev. D}\ }\textbf {\bibinfo {volume} {106}},\ \bibinfo
  {pages} {116013} (\bibinfo {year} {2022})},\ \Eprint
  {https://arxiv.org/abs/2205.06707} {arXiv:2205.06707 [hep-th]} \BibitemShut
  {NoStop}%
\bibitem [{\citenamefont {Reuter}(1998)}]{Reuter:1996cp}%
  \BibitemOpen
  \bibfield  {author} {\bibinfo {author} {\bibfnamefont {M.}~\bibnamefont
  {Reuter}},\ }\href {https://doi.org/10.1103/PhysRevD.57.971} {\bibfield
  {journal} {\bibinfo  {journal} {Phys. Rev. D}\ }\textbf {\bibinfo {volume}
  {57}},\ \bibinfo {pages} {971} (\bibinfo {year} {1998})},\ \Eprint
  {https://arxiv.org/abs/hep-th/9605030} {arXiv:hep-th/9605030} \BibitemShut
  {NoStop}%
\bibitem [{\citenamefont {Bonanno}\ \emph {et~al.}(2020)\citenamefont
  {Bonanno}, \citenamefont {Lippoldt}, \citenamefont {Percacci},\ and\
  \citenamefont {Vacca}}]{Bonanno:2019ukb}%
  \BibitemOpen
  \bibfield  {author} {\bibinfo {author} {\bibfnamefont {A.}~\bibnamefont
  {Bonanno}}, \bibinfo {author} {\bibfnamefont {S.}~\bibnamefont {Lippoldt}},
  \bibinfo {author} {\bibfnamefont {R.}~\bibnamefont {Percacci}},\ and\
  \bibinfo {author} {\bibfnamefont {G.~P.}\ \bibnamefont {Vacca}},\ }\href
  {https://doi.org/10.1140/epjc/s10052-020-7798-9} {\bibfield  {journal}
  {\bibinfo  {journal} {Eur. Phys. J. C}\ }\textbf {\bibinfo {volume} {80}},\
  \bibinfo {pages} {249} (\bibinfo {year} {2020})},\ \Eprint
  {https://arxiv.org/abs/1912.08135} {arXiv:1912.08135 [hep-th]} \BibitemShut
  {NoStop}%
\bibitem [{\citenamefont {Falls}(2021)}]{Falls:2020tmj}%
  \BibitemOpen
  \bibfield  {author} {\bibinfo {author} {\bibfnamefont {K.}~\bibnamefont
  {Falls}},\ }\href {https://doi.org/10.1140/epjc/s10052-020-08803-0}
  {\bibfield  {journal} {\bibinfo  {journal} {Eur. Phys. J. C}\ }\textbf
  {\bibinfo {volume} {81}},\ \bibinfo {pages} {121} (\bibinfo {year} {2021})},\
  \Eprint {https://arxiv.org/abs/2004.11409} {arXiv:2004.11409 [hep-th]}
  \BibitemShut {NoStop}%
\bibitem [{\citenamefont {Dietz}\ and\ \citenamefont
  {Morris}(2015)}]{Dietz:2015owa}%
  \BibitemOpen
  \bibfield  {author} {\bibinfo {author} {\bibfnamefont {J.~A.}\ \bibnamefont
  {Dietz}}\ and\ \bibinfo {author} {\bibfnamefont {T.~R.}\ \bibnamefont
  {Morris}},\ }\href {https://doi.org/10.1007/JHEP04(2015)118} {\bibfield
  {journal} {\bibinfo  {journal} {JHEP}\ }\textbf {\bibinfo {volume} {04}},\
  \bibinfo {pages} {118}},\ \Eprint {https://arxiv.org/abs/1502.07396}
  {arXiv:1502.07396 [hep-th]} \BibitemShut {NoStop}%
\bibitem [{\citenamefont {Labus}\ \emph {et~al.}(2016)\citenamefont {Labus},
  \citenamefont {Morris},\ and\ \citenamefont {Slade}}]{Labus:2016lkh}%
  \BibitemOpen
  \bibfield  {author} {\bibinfo {author} {\bibfnamefont {P.}~\bibnamefont
  {Labus}}, \bibinfo {author} {\bibfnamefont {T.~R.}\ \bibnamefont {Morris}},\
  and\ \bibinfo {author} {\bibfnamefont {Z.~H.}\ \bibnamefont {Slade}},\ }\href
  {https://doi.org/10.1103/PhysRevD.94.024007} {\bibfield  {journal} {\bibinfo
  {journal} {Phys. Rev. D}\ }\textbf {\bibinfo {volume} {94}},\ \bibinfo
  {pages} {024007} (\bibinfo {year} {2016})},\ \Eprint
  {https://arxiv.org/abs/1603.04772} {arXiv:1603.04772 [hep-th]} \BibitemShut
  {NoStop}%
\bibitem [{\citenamefont {Mandric}\ and\ \citenamefont
  {Morris}(2023)}]{Mandric:2022dte}%
  \BibitemOpen
  \bibfield  {author} {\bibinfo {author} {\bibfnamefont {V.-M.}\ \bibnamefont
  {Mandric}}\ and\ \bibinfo {author} {\bibfnamefont {T.~R.}\ \bibnamefont
  {Morris}},\ }\href {https://doi.org/10.1103/PhysRevD.107.065012} {\bibfield
  {journal} {\bibinfo  {journal} {Phys. Rev. D}\ }\textbf {\bibinfo {volume}
  {107}},\ \bibinfo {pages} {065012} (\bibinfo {year} {2023})},\ \Eprint
  {https://arxiv.org/abs/2210.00492} {arXiv:2210.00492 [hep-th]} \BibitemShut
  {NoStop}%
\bibitem [{\citenamefont {Pagani}\ and\ \citenamefont
  {Sonoda}(2024)}]{Pagani:2024lcn}%
  \BibitemOpen
  \bibfield  {author} {\bibinfo {author} {\bibfnamefont {C.}~\bibnamefont
  {Pagani}}\ and\ \bibinfo {author} {\bibfnamefont {H.}~\bibnamefont
  {Sonoda}},\ }\href@noop {} {\  (\bibinfo {year} {2024})},\ \Eprint
  {https://arxiv.org/abs/2408.03625} {arXiv:2408.03625 [hep-th]} \BibitemShut
  {NoStop}%
\bibitem [{\citenamefont {Wetterich}(2024)}]{Wetterich:2024ivi}%
  \BibitemOpen
  \bibfield  {author} {\bibinfo {author} {\bibfnamefont {C.}~\bibnamefont
  {Wetterich}},\ }\href@noop {} {\  (\bibinfo {year} {2024})},\ \Eprint
  {https://arxiv.org/abs/2403.17523} {arXiv:2403.17523 [hep-th]} \BibitemShut
  {NoStop}%
\bibitem [{\citenamefont {Wetterich}(2018{\natexlab{a}})}]{Wetterich:2016ewc}%
  \BibitemOpen
  \bibfield  {author} {\bibinfo {author} {\bibfnamefont {C.}~\bibnamefont
  {Wetterich}},\ }\href {https://doi.org/10.1016/j.nuclphysb.2018.04.020}
  {\bibfield  {journal} {\bibinfo  {journal} {Nucl. Phys. B}\ }\textbf
  {\bibinfo {volume} {931}},\ \bibinfo {pages} {262} (\bibinfo {year}
  {2018}{\natexlab{a}})},\ \Eprint {https://arxiv.org/abs/1607.02989}
  {arXiv:1607.02989 [hep-th]} \BibitemShut {NoStop}%
\bibitem [{\citenamefont {Wetterich}(2018{\natexlab{b}})}]{Wetterich:2017aoy}%
  \BibitemOpen
  \bibfield  {author} {\bibinfo {author} {\bibfnamefont {C.}~\bibnamefont
  {Wetterich}},\ }\href {https://doi.org/10.1016/j.nuclphysb.2018.07.002}
  {\bibfield  {journal} {\bibinfo  {journal} {Nucl. Phys. B}\ }\textbf
  {\bibinfo {volume} {934}},\ \bibinfo {pages} {265} (\bibinfo {year}
  {2018}{\natexlab{b}})},\ \Eprint {https://arxiv.org/abs/1710.02494}
  {arXiv:1710.02494 [hep-th]} \BibitemShut {NoStop}%
\bibitem [{\citenamefont {Wegner}(1974)}]{Wegner_1974}%
  \BibitemOpen
  \bibfield  {author} {\bibinfo {author} {\bibfnamefont {F.~J.}\ \bibnamefont
  {Wegner}},\ }\href {https://doi.org/10.1088/0022-3719/7/12/004} {\bibfield
  {journal} {\bibinfo  {journal} {Journal of Physics C: Solid State Physics}\
  }\textbf {\bibinfo {volume} {7}},\ \bibinfo {pages} {2098} (\bibinfo {year}
  {1974})}\BibitemShut {NoStop}%
\bibitem [{\citenamefont {Polchinski}(1984)}]{Polchinski1984}%
  \BibitemOpen
  \bibfield  {author} {\bibinfo {author} {\bibfnamefont {J.}~\bibnamefont
  {Polchinski}},\ }\href {https://doi.org/10.1016/0550-3213(84)90287-6}
  {\bibfield  {journal} {\bibinfo  {journal} {Nuclear Physics B}\ }\textbf
  {\bibinfo {volume} {231}},\ \bibinfo {pages} {269} (\bibinfo {year}
  {1984})}\BibitemShut {NoStop}%
\bibitem [{\citenamefont {Morris}(2000{\natexlab{a}})}]{Morris:1999px}%
  \BibitemOpen
  \bibfield  {author} {\bibinfo {author} {\bibfnamefont {T.~R.}\ \bibnamefont
  {Morris}},\ }\href {https://doi.org/10.1016/S0550-3213(99)00821-4} {\bibfield
   {journal} {\bibinfo  {journal} {Nucl. Phys. B}\ }\textbf {\bibinfo {volume}
  {573}},\ \bibinfo {pages} {97} (\bibinfo {year} {2000}{\natexlab{a}})},\
  \Eprint {https://arxiv.org/abs/hep-th/9910058} {arXiv:hep-th/9910058}
  \BibitemShut {NoStop}%
\bibitem [{\citenamefont {Morris}(2000{\natexlab{b}})}]{Morris:2000fs}%
  \BibitemOpen
  \bibfield  {author} {\bibinfo {author} {\bibfnamefont {T.~R.}\ \bibnamefont
  {Morris}},\ }\href {https://doi.org/10.1088/1126-6708/2000/12/012} {\bibfield
   {journal} {\bibinfo  {journal} {JHEP}\ }\textbf {\bibinfo {volume} {12}},\
  \bibinfo {pages} {012}},\ \Eprint {https://arxiv.org/abs/hep-th/0006064}
  {arXiv:hep-th/0006064} \BibitemShut {NoStop}%
\bibitem [{\citenamefont {Morris}\ and\ \citenamefont
  {Rosten}(2006)}]{Morris:2005tv}%
  \BibitemOpen
  \bibfield  {author} {\bibinfo {author} {\bibfnamefont {T.~R.}\ \bibnamefont
  {Morris}}\ and\ \bibinfo {author} {\bibfnamefont {O.~J.}\ \bibnamefont
  {Rosten}},\ }\href {https://doi.org/10.1103/PhysRevD.73.065003} {\bibfield
  {journal} {\bibinfo  {journal} {Phys. Rev. D}\ }\textbf {\bibinfo {volume}
  {73}},\ \bibinfo {pages} {065003} (\bibinfo {year} {2006})},\ \Eprint
  {https://arxiv.org/abs/hep-th/0508026} {arXiv:hep-th/0508026} \BibitemShut
  {NoStop}%
\bibitem [{\citenamefont {Sonoda}\ and\ \citenamefont
  {Suzuki}(2021)}]{Sonoda:2020vut}%
  \BibitemOpen
  \bibfield  {author} {\bibinfo {author} {\bibfnamefont {H.}~\bibnamefont
  {Sonoda}}\ and\ \bibinfo {author} {\bibfnamefont {H.}~\bibnamefont
  {Suzuki}},\ }\href {https://doi.org/10.1093/ptep/ptab006} {\bibfield
  {journal} {\bibinfo  {journal} {PTEP}\ }\textbf {\bibinfo {volume} {2021}},\
  \bibinfo {pages} {023B05} (\bibinfo {year} {2021})},\ \Eprint
  {https://arxiv.org/abs/2012.03568} {arXiv:2012.03568 [hep-th]} \BibitemShut
  {NoStop}%
\bibitem [{\citenamefont {Miyakawa}\ \emph {et~al.}(2022)\citenamefont
  {Miyakawa}, \citenamefont {Sonoda},\ and\ \citenamefont
  {Suzuki}}]{Miyakawa:2021wus}%
  \BibitemOpen
  \bibfield  {author} {\bibinfo {author} {\bibfnamefont {Y.}~\bibnamefont
  {Miyakawa}}, \bibinfo {author} {\bibfnamefont {H.}~\bibnamefont {Sonoda}},\
  and\ \bibinfo {author} {\bibfnamefont {H.}~\bibnamefont {Suzuki}},\ }\href
  {https://doi.org/10.1093/ptep/ptac003} {\bibfield  {journal} {\bibinfo
  {journal} {PTEP}\ }\textbf {\bibinfo {volume} {2022}},\ \bibinfo {pages}
  {023B02} (\bibinfo {year} {2022})},\ \Eprint
  {https://arxiv.org/abs/2111.15529} {arXiv:2111.15529 [hep-th]} \BibitemShut
  {NoStop}%
\bibitem [{\citenamefont {Sonoda}\ and\ \citenamefont
  {Suzuki}(2022)}]{Sonoda:2022fmk}%
  \BibitemOpen
  \bibfield  {author} {\bibinfo {author} {\bibfnamefont {H.}~\bibnamefont
  {Sonoda}}\ and\ \bibinfo {author} {\bibfnamefont {H.}~\bibnamefont
  {Suzuki}},\ }\href {https://doi.org/10.1093/ptep/ptac047} {\bibfield
  {journal} {\bibinfo  {journal} {PTEP}\ }\textbf {\bibinfo {volume} {2022}},\
  \bibinfo {pages} {053B01} (\bibinfo {year} {2022})},\ \Eprint
  {https://arxiv.org/abs/2201.04448} {arXiv:2201.04448 [hep-th]} \BibitemShut
  {NoStop}%
\bibitem [{\citenamefont {Miyakawa}\ \emph {et~al.}(2023)\citenamefont
  {Miyakawa}, \citenamefont {Sonoda},\ and\ \citenamefont
  {Suzuki}}]{Miyakawa:2023yob}%
  \BibitemOpen
  \bibfield  {author} {\bibinfo {author} {\bibfnamefont {Y.}~\bibnamefont
  {Miyakawa}}, \bibinfo {author} {\bibfnamefont {H.}~\bibnamefont {Sonoda}},\
  and\ \bibinfo {author} {\bibfnamefont {H.}~\bibnamefont {Suzuki}},\ }\href
  {https://doi.org/10.1093/ptep/ptad074} {\bibfield  {journal} {\bibinfo
  {journal} {PTEP}\ }\textbf {\bibinfo {volume} {2023}},\ \bibinfo {pages}
  {063B03} (\bibinfo {year} {2023})},\ \bibinfo {note} {[Erratum: PTEP 2024,
  039201 (2024)]},\ \Eprint {https://arxiv.org/abs/2304.14753}
  {arXiv:2304.14753 [hep-th]} \BibitemShut {NoStop}%
\bibitem [{\citenamefont {Haruna}\ and\ \citenamefont
  {Yamada}(2024)}]{Haruna:2023spq}%
  \BibitemOpen
  \bibfield  {author} {\bibinfo {author} {\bibfnamefont {J.}~\bibnamefont
  {Haruna}}\ and\ \bibinfo {author} {\bibfnamefont {M.}~\bibnamefont
  {Yamada}},\ }\href {https://doi.org/10.1007/JHEP05(2024)291} {\bibfield
  {journal} {\bibinfo  {journal} {JHEP}\ }\textbf {\bibinfo {volume} {05}},\
  \bibinfo {pages} {291}},\ \Eprint {https://arxiv.org/abs/2312.15673}
  {arXiv:2312.15673 [hep-th]} \BibitemShut {NoStop}%
\bibitem [{\citenamefont {Baldazzi}\ \emph {et~al.}(2022)\citenamefont
  {Baldazzi}, \citenamefont {Zinati},\ and\ \citenamefont
  {Falls}}]{Baldazzi:2021ydj}%
  \BibitemOpen
  \bibfield  {author} {\bibinfo {author} {\bibfnamefont {A.}~\bibnamefont
  {Baldazzi}}, \bibinfo {author} {\bibfnamefont {R.~B.~A.}\ \bibnamefont
  {Zinati}},\ and\ \bibinfo {author} {\bibfnamefont {K.}~\bibnamefont
  {Falls}},\ }\href {https://doi.org/10.21468/SciPostPhys.13.4.085} {\bibfield
  {journal} {\bibinfo  {journal} {SciPost Phys.}\ }\textbf {\bibinfo {volume}
  {13}},\ \bibinfo {pages} {085} (\bibinfo {year} {2022})},\ \Eprint
  {https://arxiv.org/abs/2105.11482} {arXiv:2105.11482 [hep-th]} \BibitemShut
  {NoStop}%
\bibitem [{\citenamefont {Ihssen}\ and\ \citenamefont
  {Pawlowski}(2023{\natexlab{b}})}]{Ihssen:2022xjv}%
  \BibitemOpen
  \bibfield  {author} {\bibinfo {author} {\bibfnamefont {F.}~\bibnamefont
  {Ihssen}}\ and\ \bibinfo {author} {\bibfnamefont {J.~M.}\ \bibnamefont
  {Pawlowski}},\ }\href {https://doi.org/10.21468/SciPostPhys.15.2.074}
  {\bibfield  {journal} {\bibinfo  {journal} {SciPost Phys.}\ }\textbf
  {\bibinfo {volume} {15}},\ \bibinfo {pages} {074} (\bibinfo {year}
  {2023}{\natexlab{b}})},\ \Eprint {https://arxiv.org/abs/2207.10057}
  {arXiv:2207.10057 [hep-th]} \BibitemShut {NoStop}%
\bibitem [{\citenamefont {Cotler}\ and\ \citenamefont
  {Rezchikov}(2022)}]{Cotler:2022fze}%
  \BibitemOpen
  \bibfield  {author} {\bibinfo {author} {\bibfnamefont {J.}~\bibnamefont
  {Cotler}}\ and\ \bibinfo {author} {\bibfnamefont {S.}~\bibnamefont
  {Rezchikov}},\ }\href@noop {} {\  (\bibinfo {year} {2022})},\ \Eprint
  {https://arxiv.org/abs/2202.11737} {arXiv:2202.11737 [hep-th]} \BibitemShut
  {NoStop}%
\bibitem [{\citenamefont {Vilkovisky}(1984)}]{Vilkovisky:1984st}%
  \BibitemOpen
  \bibfield  {author} {\bibinfo {author} {\bibfnamefont {G.~A.}\ \bibnamefont
  {Vilkovisky}},\ }\href {https://doi.org/10.1016/0550-3213(84)90228-1}
  {\bibfield  {journal} {\bibinfo  {journal} {Nucl. Phys. B}\ }\textbf
  {\bibinfo {volume} {234}},\ \bibinfo {pages} {125} (\bibinfo {year}
  {1984})}\BibitemShut {NoStop}%
\bibitem [{\citenamefont {DeWitt}(2003)}]{DeWitt:2003pm}%
  \BibitemOpen
  \bibfield  {author} {\bibinfo {author} {\bibfnamefont {B.~S.}\ \bibnamefont
  {DeWitt}},\ }\href@noop {} {\emph {\bibinfo {title} {{The global approach to
  quantum field theory. Vol. 1, 2}}}},\ Vol.\ \bibinfo {volume} {114}\
  (\bibinfo {year} {2003})\BibitemShut {NoStop}%
\bibitem [{\citenamefont {Branchina}\ \emph {et~al.}(2003)\citenamefont
  {Branchina}, \citenamefont {Meissner},\ and\ \citenamefont
  {Veneziano}}]{Branchina:2003ek}%
  \BibitemOpen
  \bibfield  {author} {\bibinfo {author} {\bibfnamefont {V.}~\bibnamefont
  {Branchina}}, \bibinfo {author} {\bibfnamefont {K.~A.}\ \bibnamefont
  {Meissner}},\ and\ \bibinfo {author} {\bibfnamefont {G.}~\bibnamefont
  {Veneziano}},\ }\href {https://doi.org/10.1016/j.physletb.2003.09.020}
  {\bibfield  {journal} {\bibinfo  {journal} {Phys. Lett. B}\ }\textbf
  {\bibinfo {volume} {574}},\ \bibinfo {pages} {319} (\bibinfo {year}
  {2003})},\ \Eprint {https://arxiv.org/abs/hep-th/0309234}
  {arXiv:hep-th/0309234} \BibitemShut {NoStop}%
\bibitem [{\citenamefont {Pawlowski}(2003)}]{Pawlowski:2003sk}%
  \BibitemOpen
  \bibfield  {author} {\bibinfo {author} {\bibfnamefont {J.~M.}\ \bibnamefont
  {Pawlowski}},\ }\href@noop {} {\  (\bibinfo {year} {2003})},\ \Eprint
  {https://arxiv.org/abs/hep-th/0310018} {arXiv:hep-th/0310018} \BibitemShut
  {NoStop}%
\bibitem [{\citenamefont {DeWitt}\ and\ \citenamefont
  {Molina-Paris}(1995)}]{DeWitt:1995cx}%
  \BibitemOpen
  \bibfield  {author} {\bibinfo {author} {\bibfnamefont {B.~S.}\ \bibnamefont
  {DeWitt}}\ and\ \bibinfo {author} {\bibfnamefont {C.}~\bibnamefont
  {Molina-Paris}},\ }in\ \href@noop {} {\emph {\bibinfo {booktitle} {{2nd
  International Sakharov Conference on Physics}}}}\ (\bibinfo {year} {1995})\
  pp.\ \bibinfo {pages} {396--408},\ \Eprint
  {https://arxiv.org/abs/hep-th/9511109} {arXiv:hep-th/9511109} \BibitemShut
  {NoStop}%
\bibitem [{\citenamefont {DeWitt}\ and\ \citenamefont
  {Molina-Paris}(1998)}]{DeWitt:1998eq}%
  \BibitemOpen
  \bibfield  {author} {\bibinfo {author} {\bibfnamefont {B.~S.}\ \bibnamefont
  {DeWitt}}\ and\ \bibinfo {author} {\bibfnamefont {C.}~\bibnamefont
  {Molina-Paris}},\ }\href {https://doi.org/10.1142/S0217732398002631}
  {\bibfield  {journal} {\bibinfo  {journal} {Mod. Phys. Lett. A}\ }\textbf
  {\bibinfo {volume} {13}},\ \bibinfo {pages} {2475} (\bibinfo {year}
  {1998})},\ \Eprint {https://arxiv.org/abs/hep-th/9808163}
  {arXiv:hep-th/9808163} \BibitemShut {NoStop}%
\bibitem [{\citenamefont {Capri}\ \emph {et~al.}(2015)\citenamefont {Capri},
  \citenamefont {Dudal}, \citenamefont {Fiorentini}, \citenamefont {Guimaraes},
  \citenamefont {Justo}, \citenamefont {Pereira}, \citenamefont {Mintz},
  \citenamefont {Palhares}, \citenamefont {Sobreiro},\ and\ \citenamefont
  {Sorella}}]{Capri:2015ixa}%
  \BibitemOpen
  \bibfield  {author} {\bibinfo {author} {\bibfnamefont {M.~A.~L.}\
  \bibnamefont {Capri}}, \bibinfo {author} {\bibfnamefont {D.}~\bibnamefont
  {Dudal}}, \bibinfo {author} {\bibfnamefont {D.}~\bibnamefont {Fiorentini}},
  \bibinfo {author} {\bibfnamefont {M.~S.}\ \bibnamefont {Guimaraes}}, \bibinfo
  {author} {\bibfnamefont {I.~F.}\ \bibnamefont {Justo}}, \bibinfo {author}
  {\bibfnamefont {A.~D.}\ \bibnamefont {Pereira}}, \bibinfo {author}
  {\bibfnamefont {B.~W.}\ \bibnamefont {Mintz}}, \bibinfo {author}
  {\bibfnamefont {L.~F.}\ \bibnamefont {Palhares}}, \bibinfo {author}
  {\bibfnamefont {R.~F.}\ \bibnamefont {Sobreiro}},\ and\ \bibinfo {author}
  {\bibfnamefont {S.~P.}\ \bibnamefont {Sorella}},\ }\href
  {https://doi.org/10.1103/PhysRevD.92.045039} {\bibfield  {journal} {\bibinfo
  {journal} {Phys. Rev. D}\ }\textbf {\bibinfo {volume} {92}},\ \bibinfo
  {pages} {045039} (\bibinfo {year} {2015})},\ \Eprint
  {https://arxiv.org/abs/1506.06995} {arXiv:1506.06995 [hep-th]} \BibitemShut
  {NoStop}%
\bibitem [{\citenamefont {Falls}(2025)}]{Falls:2025tid}%
  \BibitemOpen
  \bibfield  {author} {\bibinfo {author} {\bibfnamefont {K.}~\bibnamefont
  {Falls}},\ }\href@noop {} {\  (\bibinfo {year} {2025})},\ \Eprint
  {https://arxiv.org/abs/2503.05869} {arXiv:2503.05869 [hep-th]} \BibitemShut
  {NoStop}%
\bibitem [{\citenamefont {Donkin}\ and\ \citenamefont
  {Pawlowski}(2012)}]{Donkin:2012ud}%
  \BibitemOpen
  \bibfield  {author} {\bibinfo {author} {\bibfnamefont {I.}~\bibnamefont
  {Donkin}}\ and\ \bibinfo {author} {\bibfnamefont {J.~M.}\ \bibnamefont
  {Pawlowski}},\ }\href@noop {} {\  (\bibinfo {year} {2012})},\ \Eprint
  {https://arxiv.org/abs/1203.4207} {arXiv:1203.4207 [hep-th]} \BibitemShut
  {NoStop}%
\bibitem [{\citenamefont {Frohlich}\ \emph {et~al.}(1981)\citenamefont
  {Frohlich}, \citenamefont {Morchio},\ and\ \citenamefont
  {Strocchi}}]{Frohlich:1981yi}%
  \BibitemOpen
  \bibfield  {author} {\bibinfo {author} {\bibfnamefont {J.}~\bibnamefont
  {Frohlich}}, \bibinfo {author} {\bibfnamefont {G.}~\bibnamefont {Morchio}},\
  and\ \bibinfo {author} {\bibfnamefont {F.}~\bibnamefont {Strocchi}},\ }\href
  {https://doi.org/10.1016/0550-3213(81)90448-X} {\bibfield  {journal}
  {\bibinfo  {journal} {Nucl. Phys. B}\ }\textbf {\bibinfo {volume} {190}},\
  \bibinfo {pages} {553} (\bibinfo {year} {1981})}\BibitemShut {NoStop}%
\bibitem [{\citenamefont {Egger}\ \emph {et~al.}(2017)\citenamefont {Egger},
  \citenamefont {Maas},\ and\ \citenamefont {Sondenheimer}}]{Egger:2017tkd}%
  \BibitemOpen
  \bibfield  {author} {\bibinfo {author} {\bibfnamefont {L.}~\bibnamefont
  {Egger}}, \bibinfo {author} {\bibfnamefont {A.}~\bibnamefont {Maas}},\ and\
  \bibinfo {author} {\bibfnamefont {R.}~\bibnamefont {Sondenheimer}},\ }\href
  {https://doi.org/10.1142/S0217732317502121} {\bibfield  {journal} {\bibinfo
  {journal} {Mod. Phys. Lett. A}\ }\textbf {\bibinfo {volume} {32}},\ \bibinfo
  {pages} {1750212} (\bibinfo {year} {2017})},\ \Eprint
  {https://arxiv.org/abs/1701.02881} {arXiv:1701.02881 [hep-ph]} \BibitemShut
  {NoStop}%
\bibitem [{\citenamefont {Maas}\ \emph {et~al.}(2019)\citenamefont {Maas},
  \citenamefont {Sondenheimer},\ and\ \citenamefont {T\"orek}}]{Maas:2017xzh}%
  \BibitemOpen
  \bibfield  {author} {\bibinfo {author} {\bibfnamefont {A.}~\bibnamefont
  {Maas}}, \bibinfo {author} {\bibfnamefont {R.}~\bibnamefont {Sondenheimer}},\
  and\ \bibinfo {author} {\bibfnamefont {P.}~\bibnamefont {T\"orek}},\ }\href
  {https://doi.org/10.1016/j.aop.2019.01.010} {\bibfield  {journal} {\bibinfo
  {journal} {Annals Phys.}\ }\textbf {\bibinfo {volume} {402}},\ \bibinfo
  {pages} {18} (\bibinfo {year} {2019})},\ \Eprint
  {https://arxiv.org/abs/1709.07477} {arXiv:1709.07477 [hep-ph]} \BibitemShut
  {NoStop}%
\bibitem [{\citenamefont {Sondenheimer}(2020)}]{Sondenheimer:2019idq}%
  \BibitemOpen
  \bibfield  {author} {\bibinfo {author} {\bibfnamefont {R.}~\bibnamefont
  {Sondenheimer}},\ }\href {https://doi.org/10.1103/PhysRevD.101.056006}
  {\bibfield  {journal} {\bibinfo  {journal} {Phys. Rev. D}\ }\textbf {\bibinfo
  {volume} {101}},\ \bibinfo {pages} {056006} (\bibinfo {year} {2020})},\
  \Eprint {https://arxiv.org/abs/1912.08680} {arXiv:1912.08680 [hep-th]}
  \BibitemShut {NoStop}%
\bibitem [{\citenamefont {Maas}\ and\ \citenamefont
  {Sondenheimer}(2020)}]{Maas:2020kda}%
  \BibitemOpen
  \bibfield  {author} {\bibinfo {author} {\bibfnamefont {A.}~\bibnamefont
  {Maas}}\ and\ \bibinfo {author} {\bibfnamefont {R.}~\bibnamefont
  {Sondenheimer}},\ }\href {https://doi.org/10.1103/PhysRevD.102.113001}
  {\bibfield  {journal} {\bibinfo  {journal} {Phys. Rev. D}\ }\textbf {\bibinfo
  {volume} {102}},\ \bibinfo {pages} {113001} (\bibinfo {year} {2020})},\
  \Eprint {https://arxiv.org/abs/2009.06671} {arXiv:2009.06671 [hep-ph]}
  \BibitemShut {NoStop}%
\bibitem [{\citenamefont {Ihssen}\ \emph {et~al.}(2024)\citenamefont {Ihssen},
  \citenamefont {Pawlowski}, \citenamefont {Sattler},\ and\ \citenamefont
  {Wink}}]{Ihssen:2024miv}%
  \BibitemOpen
  \bibfield  {author} {\bibinfo {author} {\bibfnamefont {F.}~\bibnamefont
  {Ihssen}}, \bibinfo {author} {\bibfnamefont {J.~M.}\ \bibnamefont
  {Pawlowski}}, \bibinfo {author} {\bibfnamefont {F.~R.}\ \bibnamefont
  {Sattler}},\ and\ \bibinfo {author} {\bibfnamefont {N.}~\bibnamefont
  {Wink}},\ }\href@noop {} {\  (\bibinfo {year} {2024})},\ \Eprint
  {https://arxiv.org/abs/2408.08413} {arXiv:2408.08413 [hep-ph]} \BibitemShut
  {NoStop}%
\bibitem [{\citenamefont {Ferreira}\ and\ \citenamefont
  {Papavassiliou}(2025)}]{Ferreira:2025anh}%
  \BibitemOpen
  \bibfield  {author} {\bibinfo {author} {\bibfnamefont {M.~N.}\ \bibnamefont
  {Ferreira}}\ and\ \bibinfo {author} {\bibfnamefont {J.}~\bibnamefont
  {Papavassiliou}},\ }\href@noop {} {\  (\bibinfo {year} {2025})},\ \Eprint
  {https://arxiv.org/abs/2501.01080} {arXiv:2501.01080 [hep-ph]} \BibitemShut
  {NoStop}%
\bibitem [{\citenamefont {Abbott}(1981)}]{Abbott:1980hw}%
  \BibitemOpen
  \bibfield  {author} {\bibinfo {author} {\bibfnamefont {L.~F.}\ \bibnamefont
  {Abbott}},\ }\href {https://doi.org/10.1016/0550-3213(81)90371-0} {\bibfield
  {journal} {\bibinfo  {journal} {Nucl. Phys. B}\ }\textbf {\bibinfo {volume}
  {185}},\ \bibinfo {pages} {189} (\bibinfo {year} {1981})}\BibitemShut
  {NoStop}%
\end{thebibliography}%
	
\end{document}